\documentclass[a4paper,11pt]{article}
\pdfoutput=1

\usepackage[utf8]{inputenc}
\usepackage[T1]{fontenc}
\usepackage[english]{babel}
\usepackage{layout}
\usepackage{setspace}
\usepackage{lmodern}
\usepackage{enumitem}
\usepackage{soul}
\usepackage{eurosym}

\usepackage{verbatim}
\usepackage{moreverb}

\usepackage{sectsty}

\usepackage{mathrsfs}
\usepackage{cancel}

\usepackage{makeidx}
\usepackage{multirow}
\usepackage{stmaryrd}

\usepackage{docstyle}

\renewcommand{\bar}[1]{\overline{\mathstrut #1}}
\renewcommand{\vec}[1]{\overrightarrow{\mathstrut #1}}

\title{\textcolor{mycolor}{Large Star/Rose Extra Dimension\\with Small Leaves/Petals}}

\author{Florian Nortier}

\affiliation{Université Paris-Saclay, CNRS/IN2P3, IJCLab, 91405 Orsay, France}

\emailAdd{florian.nortier@protonmail.com}

\abstract{In this paper, we propose to compactify a single large extra dimension (LED) on a
star/rose graph with a large number of identical leaves/petals. The 5D Planck scale can
be chosen to be $\Lambda_P^{(5)} \sim \mathcal{O}(1)$ TeV which can provide a path to solve the gauge hierarchy
problem. The leaf/petal length scale is of $\mathcal{O}(1/\Lambda_{EW})$, where $\Lambda_{EW} \sim 100$ GeV is the
weak scale, without the large geometrical hierarchy of the traditional LED models to
stabilize. The 4D fields of the SM are localized on a 3-brane at the central vertex of the
star/rose graph. We predict a tower of feebly coupled weak scale Kaluza-Klein (KK)
gravitons below a regime of strongly coupled gravitational phenomena above the TeV
scale. Moreover, we reformulate in our setup the LED mechanism to generate light Dirac
neutrinos, where the right-handed neutrinos are KK-modes of gauge singlet fermions
propagating in the bulk. A large number of KK-gravitons and KK-neutrinos interact
only gravitationally, and thus constitute a hidden sector.}

\keywords{Beyond the Standard Model; Large Extra Dimensions; Braneworlds; Quantum
Graphs; Neutrino Physics; Hidden Sector.}

\begin{document}

\maketitle
\flushbottom

\section{Introduction}
The large gauge hierarchy between the 4D Planck scale,
\begin{equation}
\Lambda_P^{(4)} = \sqrt{\frac{1}{8 \pi G_N^{(4)}}} \simeq 2.4 \times 10^{18} \  \mathrm{GeV},
\end{equation}
where $G_N^{(4)}$ is the 4D gravitational Newton constant, and the electroweak (EW) scale
$\Lambda_{EW} \sim 100$~GeV questions the naturalness of the discovered light Higgs boson of the
standard model (SM) of particle physics coupled to gravity \cite{Wilson:1970ag,Susskind:1978ms,tHooft:1979rat}. This gauge hierarchy
problem has motivated many beyond the SM-scenarios for the last decades. A
possibility to solve this issue is to embed the SM into a theory where the true
gravity scale, i.e. the scale where gravitational self-interactions become strong and
new degrees of freedom are expected, is not at the 4D Planck scale but in the
$1-10$~TeV range. However, this does not solve the question of how an ultraviolet (UV)
completion of quantum gravity is achieved.\\

If there are flat extra dimensions of space (EDS's) on a $q$-dimensional compact space $\mathcal{C}_q$ of volume $\mathcal{V}_q$, with a factorizable spacetime geometry $\mathcal{M}_4 \times \mathcal{C}_q$ where $\mathcal{M}_4$ is the 4D Minkowski spacetime, the 4D Planck scale $\Lambda_P^{(4)}$ is just an effective scale given by the relation
\begin{equation}
\left[\Lambda_P^{(4)}\right]^2 = \left[ \Lambda_P^{(4+q)} \right]^{q+2} \, \mathcal{V}_q \, ,
\label{ADD_formula}
\end{equation}
involving the $(4+q)$D Planck scale
\begin{equation}
\Lambda_P^{(4+q)} = \left[ \dfrac{1}{8 \pi G^{(4+q)}_N} \right]^{1/(q+2)},
\end{equation}
where $G_N^{(4+q)}$ is the $(4+q)$D gravitational Newton constant. $\Lambda_P^{(4+q)}$ is the real scale at which gravity becomes strongly coupled\footnote{The idea of reducing the gravity scale with an EDS in string theory is due to Witten in Ref.~\cite{Witten:1996mz}.}, so it is the true cut-off of the quantum field theory (QFT)\footnote{To be more precise, if the UV-completion is perturbative, the UV-gravitational degrees of freedom appear at a scale $\Lambda_{UV}<\Lambda_P^{(4+q)}$. The cut-off of the effective field theory (EFT) is $\Lambda_{UV}$ and not $\Lambda_P^{(4+q)}$. For example, in perturbative string theory, the string scale is lower than the higher-dimensional Planck scale. To simplify the discussion, we ignore this possibility here.}. In 1998, Arkani-Hamed, Dimopoulos and Dvali (ADD) proposed in Ref.~\cite{ArkaniHamed:1998rs} to use this result to solve the gauge hierarchy problem if $\Lambda_P^{(4+q)}  \sim \mathcal{O}(1)$ TeV with a large compactified volume $\mathcal{V}_q$. In ADD models, the SM-fields must be localized on a 3-brane, in contrast to gravity which is a property of $(4+q)D$ spacetime in general relativity. At large distances between two test masses on the 3-brane, gravity appears as a feebly coupled theory, because gravitational fluxes spread into the large volume $\mathcal{V}_q$ of the bulk. Eventually \cite{ArkaniHamed:1998nn}, it was realized that the Kaluza-Klein (KK) modes of fields propagating into this large volume $\mathcal{V}_q$ have couplings to the SM-fields suppressed by $\sqrt{\mathcal{V}_q}$. One can thus build natural models of feebly interacting particles, i.e. particles which have a tiny coupling constant with the SM, like right-handed neutrinos \cite{Dienes:1998sb,ArkaniHamed:1998vp,Dvali:1999cn}, axions \cite{Chang:1999si,Dienes:1999gw}, dark photons \cite{ArkaniHamed:1998nn}, etc. In Ref.~\cite{ArkaniHamed:1998nn}, ADD proposed a simple toroidal compactification $\mathcal{C}_q = \left(\mathcal{R}_1\right)^q$, where $\mathcal{R}_1$ is the circle of radius $R$, and $\mathcal{V}_q = (2 \pi R)^q$. The bulk fields, like the graviton, generate a tower of KK-modes with a uniform mass gap of $1/R$. For a benchmark value $\Lambda_P^{(4+q)} = 1$ TeV, the $(4+q)$D Planck length is $\ell_P^{(4+q)} = 1/ \Lambda_P^{(4+q)} \simeq 2 \times 10^{-19} \ \mathrm{m}$, and one gets Tab.~\ref{table}.

\begin{table}[h]
\begin{center}
\begin{tabular}{c||c|c|c}
$q$ & $R$ (m) & $R/\ell_P^{(4+q)}$ & $M_{KK}$ (eV)  \\
\hline \hline
1 & $2 \times 10^{11}$ & $9 \times 10^{29}$ & $1 \times 10^{-18}$ \\ 
2 & $8 \times 10^{-5}$ & $4 \times 10^{14}$ & $3 \times 10^{-3}$ \\
4 & $2 \times 10^{-12}$ & $8 \times 10^{6}$ & $1 \times 10^5$ \\
6 & $4 \times 10^{-15}$ & $2 \times 10^{4}$ & $5 \times 10^7$
\end{tabular}
\caption{$R$, $R/\ell_P^{(4+q)}$ and $M_{KK}$ as a function of $q$ for $\Lambda_P^{(4+q)} = 1 \ \mathrm{TeV}$.}
\label{table}
\end{center}
\end{table}

Motivated by UV-completions in superstring/M-theory \cite{Antoniadis:1998ig,Benakli:1998pw} requiring 10/11 spacetime dimensions, most of the efforts concentrated on $q \leq 7$. The compactification radius $R$ must be stabilized at a large value compared to $\ell_P^{(4+q)}$, which reintroduces a geometrical hierarchy \cite{ArkaniHamed:1998nn} with a low KK-mass gap: too light KK-gravitons are constrained by astrophysics, cosmology and collider physics. When one probes gravitational Newton's law at large [small] distances with respect to $R$, gravity appears 4D [$(4+q)$D]. The case $q=1$ is excluded because it leads to a modification of 4D gravitational Newton's law at the scale of the solar system. ADD's proposal is thus often associated with the large extra dimensions (LED's) paradigm, and is just a reformulation of the gauge hierarchy problem. In the literature, there are interesting propositions to stabilize such large compactification radii \cite{ArkaniHamed:1998kx, ArkaniHamed:1999dz, Mazumdar:2001ya, Carroll:2001ih, Albrecht:2001cp, Antoniadis:2002gw, Peloso:2003nv}.\\

Most of the works in the literature focus on compactifications on manifolds or orbifolds, the latter are obtained by modding out a manifold with a discrete isometry. The reason is that string theories are well-defined on these geometries. However, as noticed in Ref.~\cite{Hebecker:2001jb}, working with an EFT allows us to consider more general geometries which are not obtained as coset spaces, as long as the resulting QFT is consistent. The goal of the present work is to discuss such nonstandard compactification geometries, which do not introduce the geometrical hierarchy problem in traditional ADD-like models. In 2005, Kim proposed in Ref.~\cite{Kim:2005aa} to realize ADD's idea by compactifying a LED on a 1D singular variety: a metric graph\footnote{Metric graphs have interesting applications in physics, chemistry and mathematics (cf. Ref.~\cite{Kuchment2002} for a short review). A 2D QFT on a star graph background was developed in Refs.~\cite{Bellazzini:2006jb, Bellazzini:2006kh, Bellazzini:2008mn,Bellazzini:2008cs}. The reader can find a mathematical introduction to the spectral analysis of differential operators defined on metric graphs, the so-called quantum graphs, in Ref.~\cite{Kuchment2014}.}, like a star or a rose with respectively $N$ leaves/petals of equal length/circumference $\ell$. In Ref.~\cite{Kim:2005aa}, it was shown that, for large $N$, one can build a phenomenologically viable model with only a single LED which gives sizable submillimeter deviations from the Newtonian law of gravity in tabletop experiments. The KK-mass scale is $M_{KK} = 1/\ell \sim \mathcal{O}(10-100)$ meV. Here, we want to push the concept further, and we take $\ell$ close to $\ell_P^{(5)}$ for large $N$, so $M_{KK} = 1/\ell \sim \mathcal{O}(0.1-1)$ TeV which does not reintroduce a scale hierarchy and evade all constrains on traditional ADD models (with a compactification on a low dimensional torus) from submillimeter tests of Newtonian gravity, astrophysics and cosmology. The integer $N$ is radiatively stable, so the scenario solves completely the naturalness problem of the Higgs mass under the hypothesis of an exact global permutation symmetry between the leaves/petals. Ref.~\cite{Kim:2005aa} gives no information on the way to embed the SM-fields into the proposed geometry. Are they bulk or brane-localized fields? In this work, we will see that the SM-fields must be localized on a 3-brane, and we find it particularly interesting to localize them on the junction (central vertex) of the star/rose graph.\\

In the context of the compactification of an EDS on a metric graph, the star graph is the most popular \cite{Kim:2005aa, Cacciapaglia:2006tg, Bechinger:2009qk, Abel:2010kw, Law:2010pv, Fonseca:2019aux}, mainly because, with AdS$_5$ leaves, these effective 5D braneworlds capture the low energy behavior of models with warped throats, arising in flux compactification in type IIB superstring theory \cite{Verlinde:1999fy, Klebanov:2000hb, Giddings:2001yu, Dimopoulos:2001ui, Dimopoulos:2001qd, Kachru:2003aw, Cacciapaglia:2006tg}, when one integrates out the modes associated to the transverse dimensions of the throats. In this work, we study an EDS compactified on a flat star/rose graph with identical leaves/petals by adopting a bottom-up approach: we are completely agnostic about the origin of this curious geometry in a UV-theory.\\

This article is organized as follows. In Section~\ref{spacetime_geom}, we give the definitions of a star and a rose graph, we discuss the symmetries of the two 5D spacetimes we consider, and we introduce the elements of the distribution theory we need to define a field theory on them.\\

Before building particle physics models with a star/rose EDS, it is useful to have an idea of the mass spectra and degeneracy of the KK-modes of 5D bosons and fermions. For that purpose, we study a real scalar field in Section~\ref{KG_field} and a spin 1/2 Dirac fermion in Section~\ref{Dirac_field}. We analyze carefully the different possibilities for the junction conditions of the 5D fields, in particular the consequences of the continuity hypothesis of the fields at the central vertex of the star/rose graph.\\

In Section~\ref{ADD_star_rose}, we propose a model to reduce the gravity scale to the TeV-scale with a large compactified volume, but with EW and KK-scales which coincide. The SM-fields are localized on the 3-brane at the central vertex of the star/rose, and we compute their couplings to spinless KK-gravitons. We find that the results are very different from standard ADD models in the literature, due to the very specific features of the star/rose graph with identical leaves/petals. We also discuss briefly what kind of physics is expected in the Planckian and trans-Planckian regime of the model, the possibility of a hidden sector made of KK-gravitons and of a dark matter candidate: a stable black hole remnant \cite{Koch:2005ks, Dvali:2010gv, Bellagamba:2012wz, Alberghi:2013hca}, the Planckion \cite{Treder:1985kb, Dvali:2016ovn}.\\

In Section~\ref{Dirac_Neutrinos}, we revisit the models of Refs.~\cite{Dienes:1998sb, ArkaniHamed:1998vp, Dvali:1999cn}, which generate small Dirac neutrino masses with right-handed neutrinos identified with the zero-modes of gauge singlet fermions propagating in a large compactified volume, by adapting this idea to our spacetime geometries. We consider a toy model with only one generation of neutrinos. We use alternatively the zero-mode approximation and the exact treatment concerning the brane-localized Yukawa coupling between the SM-Higgs field with the 5D neutrino and the 4D left-handed neutrino of the SM-particle content. For this exact treatment of a brane-localized Yukawa interaction, we use the 5D method that we developed in Ref.~\cite{Angelescu:2019viv, Nortier:2020xms} with other authors. We find that a large number of KK-neutrinos are sterile and are part of the hidden sector of the proposed models.\\

We conclude and propose some perspectives in Section~\ref{conclusion_star_rose}. In Appendix~\ref{conventions}, we give our conventions for the 5D Minkowski metric, the Dirac matrices and spinors.

\section{Star/Rose Extra Dimension}
\label{spacetime_geom}
We want to study a model with a flat EDS compactified on a 1D singular variety called a metric graph. A quantum graph is by definition a metric graph equipped with a Hamiltonian and vertex conditions which guarantee that this Hamiltonian is self-adjoint. A field theory defined on a metric graph is thus a quantum graph problem. Indeed, once the action is given, Hamilton's principle give the Euler-Lagrange equations of the fields on the bonds and the vertex conditions. The reader is referred to Chapter 1 of Ref.~\cite{Kuchment2014} for basic definitions, vocabulary and properties of metric and quantum graphs which we will use in what follows. We choose to study only the cases of a star and a rose graph.

\subsection{Star \& Rose Graphs}
In this subsection, we give the definitions of a star and a rose graph.

\begin{figure}[h]
\begin{center}
\includegraphics[height=7cm]{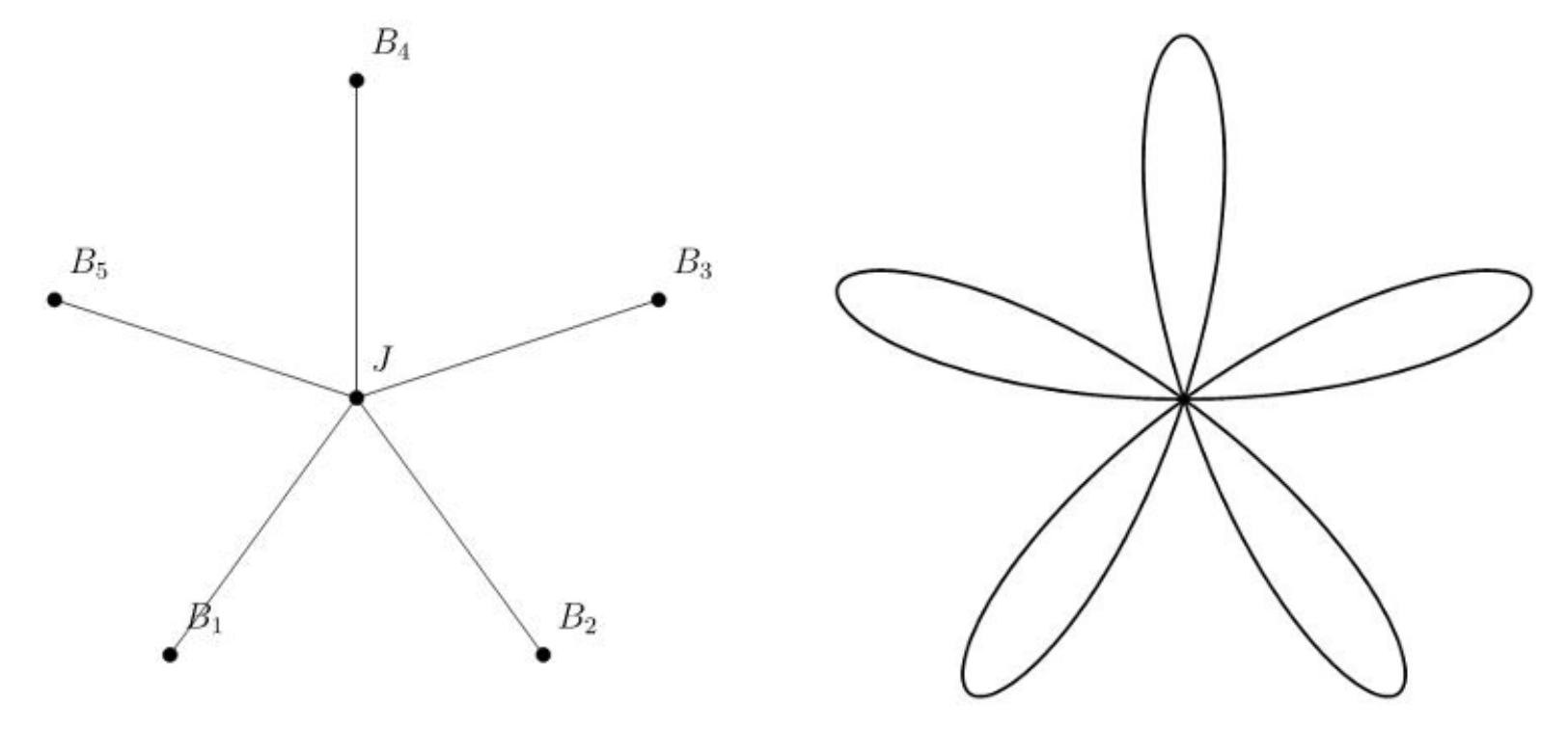}
\end{center}
\caption{Embeddings of a 5-star $\mathcal{S}_5$ (on the left) and of a 5-rose $\mathcal{R}_5$ (on the right) in $\mathbb{R}^2$.}
\label{star_rose_graph}
\end{figure}

\paragraph{$N$-star --}
The $N$-star $\mathcal{S}_N$ (cf. Fig.~\ref{star_rose_graph}) is defined as the flat equilateral star graph with $N$ bonds directed from 1 vertex of degree $N$ to $N$ vertices of degree 1. It is a flat 1D space of volume $L = N \ell$ obtained by gluing $N$ intervals (the leaves) of length $\ell$ at a common boundary $J$ (the junction). The $i^\text{th}$ leaf ends at the opposite side of the junction $J$: the boundary $B_i$. $\mathcal{S}_N$ is symmetric under the group $\Sigma_N$, which is the set of all permutations of the $N$ leaves. For example, $\mathcal{S}_1$ is the interval of length $\ell$, $\mathcal{S}_2$ the interval of length $2\ell$ symmetric under a reflection ($\Sigma_2 \simeq \mathbb{Z}_2$) with respect to the midpoint $J$, and $\mathcal{S}_3$ is a claw. The couple of coordinates $(y, i) \in [0,\ell] \times \llbracket 1, N \rrbracket$ are assigned to every point of the $i^\text{th}$ leaf with the identification:
\begin{equation}
\forall (i, j) \in \llbracket 1, N \rrbracket^2, \ i \neq j, \ (0, i) \sim (0, j) \, .
\end{equation}

\paragraph{$N$-rose --}
The $N$-rose $\mathcal{R}_N$ (cf. Fig.~\ref{star_rose_graph}) is defined as the flat equilateral rose graph (also called rhodonea or bouquet of circles), with $N$ directed loops (1 vertex of degree $2N$). It is a flat 1D space of volume $L=N\ell$ obtained by gluing the boundaries of $N$ intervals (the petals), of radius $R$ and circumference $\ell = 2 \pi R$, at a single point $V$ (the vertex/junction). $\mathcal{R}_N$ is symmetric under the group $\Sigma_N$, which is the set of all permutations of the $N$ petals. For example, $\mathcal{R}_1$ is a circle, $\mathcal{R}_2$ a lemniscat, $\mathcal{R}_3$ a trifolium, and $\mathcal{R}_4$ a quadrifolium. The couple of coordinates $(y, i) \in [0, \ell] \times \llbracket 1, N \rrbracket$ is assigned to every point of the $i^\text{th}$ petal, with the identifications:
\begin{equation}
\forall i \in \llbracket 1, N \rrbracket, \ (0, i) \sim (\ell, i) \, ,
\end{equation}
and
\begin{equation}
\forall (i, j) \in \llbracket 1, N \rrbracket^2, \ i \neq j, \, (0, i) \sim (0, j) \, .
\end{equation}

\subsection{Spacetime Geometry}
\label{Spacetime_Geometry}
We want to study a field theory on the flat factorizable geometry $\mathcal{M}_4 \times \mathcal{K}_N$, with $\mathcal{K}_N = \mathcal{S}_N$ or $\mathcal{R}_N$. The coordinates can be split as $z^M_i = (x^\mu, y, i)$, where $x^\mu$ are the coordinates of $\mathcal{M}_4$. One has $M  \in \llbracket 0, 4 \rrbracket$, and $\mu \in \llbracket 0, 3 \rrbracket$. Each leaf/petal is equipped with a 5D flat metric $\eta_{i,MN}$. We stress that one has to consider the whole star/rose graph as only one EDS. By construction, the 5D Lorentz-Poincaré symmetries are preserved locally outside the vertices. The junction $J/V$ and the boundaries $B_i$ break explicitly the 5D Lorentz-Poincaré symmetries to the 4D ones at their positions. The timelike hyperplanes at the vertices have thus the properties of 3-branes. The 3-branes at the boundaries are called $B_i$-branes, and the 3-brane at the junction is called $J/V$-brane for $\mathcal{K}_N = \mathcal{S}_N/\mathcal{R}_N$.\\

We adopt a bottom-up approach, i.e. we stay agnostic about the UV-completion of this kind of geometries, in particular the microscopic description of the 3-branes at the vertices. For instance, there are various arguments that gravity implies the existence of a minimal length scale in Nature of the order of the fundamental Planck length (cf. Ref.~\cite{Hossenfelder:2012jw} for a review). Therefore, in a UV-completion including gravity, the singular behavior of the vertices should be regularized. In Ref.~\cite{EXNER200577}, it is shown that the spectrum of a quantum graph can arise in the thin limit of a ``graph-like manifold''. Therefore, the metric graph structure of our EDS could emerge from a UV-completion where the internal space is a $q$D graph-like manifold with $q-1$ transverse dimensions of 5D Planck size, and where the vertex at the junction is regularized with a protrusion \cite{Cacciapaglia:2006tg}. After integrating out the modes associated to these $q-1$ compactified transverse dimensions, one is left with a 5D EFT with only one EDS compactified on a metric graph.\\

In this bottom-up approach, one can consider all the possibilities of field localization allowed by the spacetime symmetries. One can thus take 5D fields which propagate only into some leaves/petals as in Refs.~\cite{Cacciapaglia:2006tg, Bechinger:2009qk, Abel:2010kw, Law:2010pv, Fonseca:2019aux}, or 5D fields which propagate into the whole star/rose graph \cite{Kim:2005aa}. Moreover, 4D fields and brane-localized kinetic and/or interaction terms for the bulk fields can be strictly localized on the branes at the vertex positions. The spacetime symmetries allow 5D fields which propagate into several leaves/petals to be discontinuous at the junction and thus multivalued on this brane. It is not a problem as we consider an EFT: it just means that the field vary sharply across the brane on a distance which is less than the range of validity of the EFT. In this picture, the value of a 5D field at the point $(x^\mu, 0, i)$ is the one at the neighborhood but outside the core of the $J/V$-brane: one should consider the point $(x^\mu, 0,i)$ of the graph as $(x^\mu, 0^+, i)$. The continuity or discontinuity of a 5D field across the junction depends on the microscopic description of the vertex and is a model building hypothesis in a bottom-up approach. We stress that the distributional partial derivatives $\partial_y^n$ of a discontinuous field involve singular distributions. As the action must be finite, it is important to write operators involving product of field partial derivatives in a way which is well-defined in distribution theory (cf. Subsection~\ref{KG_field_star} for an example).\\

One can notice that there is another equivalent way to build a 5D field theory on a star/rose background. One can begin with the traditional circle compactification $\mathcal{M}_4 \times \mathcal{R}_1$ with the coordinate along the EDS $y \in [0, 2\pi R)$. Consider $N$ equidistant 3-branes along the circle, i.e. at $y_i = 2 \pi (i-1)/N$, $i \in \llbracket 1, N \rrbracket$. One can then perform the identifications: $y_i \sim y_{i+1}$ to obtain a $N$-rose. The Lagrangian has to be equal on these branes: $\mathcal{L}(y_i)=\mathcal{L}(y_{i+1})$. If a 4D field $\phi_i$ is localized on the 3-brane at $y=y_i$, then a copy of this field must be localized on all the other 3-branes. A sufficient condition is $\phi_i (x^\mu)=\phi_{i+1}(x^\mu)$. For a 5D field $\Phi$ which is continuous across all the branes, a sufficient condition is $\Phi(x^\mu, y_i)=\Phi(x^\mu, y_{i+1})$. If $\Phi$ is discontinuous across the branes, the precise value of the field at the brane position is irrelevant (remember that we work with fields as distributions). The only consistency requirement is that the Noether currents are conserved across the branes. Now, to obtain a $N$-star, one can consider the previous identifications to obtain a $N$-rose from the circle, and then notice that each petal is symmetric under a $\mathbb{Z}_2$ reflection with respect to the axis going through the center of the petal and the vertex $V$. By modding out each petal by a $\mathbb{Z}_2$ symmetry, one gets a $N$-star with the boundaries $B_i$ (fixed points) at the middle of the intervals $[y_i, y_{i+1}]$ on the circle. Therefore, instead of working on a star/rose graph, one can instead work on a circle with the previous identifications: the two pictures are physically equivalent. In the following, we choose to use the metric graph description.\\

\subsection{Distribution Theory on a Star/Rose Graph}
Before building any concrete model, it is important to define an EFT with fields of different spins on this kind of backgrounds. The authors of Ref.~\cite{Cacciapaglia:2006tg} analyzed a Klein-Gordon field, a Dirac field, a Maxwell field and Einsteinian gravity propagating in an EDS compactified on a star with $N$ leaves of different lengths. For that purpose, they define a copy of the same 5D field on each leaf. The copies are connected at the junction of the star through brane-localized interactions and the continuity of the metric. Of course, different 5D fields on each leaf are not equivalent to only one 5D field defined on the whole star graph. In order to recover only one zero-mode propagating on the whole star, they add brane-localized mass terms and take the limit of infinite masses such that $N-1$ zero-modes decouple from the EFT. However, the meaning of an infinite brane-localized mass term is not clear when the cut-off of the EFT is not so far above the KK-scale $M_{KK} = 1/\ell$. That is why we choose in this work the more straightforward approach of Refs.~\cite{Kim:2005aa, Fujimoto:2019fzb} where a 5D field is defined on the whole metric graph from the start. For that purpose, one needs a distribution theory on the star/rose graph allowing to define a Lagrangian with possible field discontinuities at the junction.\\

In this subsection, we consider distributions acting on functions defined on a star/rose graph by ignoring the 4D transverse dimensions, but it is straightforward to generalize our discussion to the case of 5D fields. To be general, we allow the test functions to be discontinuous at the junction. The usual Schwartz's distribution theory \cite{Schwartz1, Schwartz2} is thus not suitable and one should consider instead a generalization on metric graphs of Kurasov's distribution theory \cite{KURASOV1996297}. To our knowledge, such a generalization on metric graphs was not considered in the literature. We will thus define in this subsection the objects we need for our study.\\

\paragraph{Function on $\mathcal{K}_N$ --}
A complex function $f$ on the metric graph $\mathcal{K}_N$ is defined as:
\begin{equation}
f: \left\{
\begin{array}{ccc}
[0, \ell] \times \llbracket 1, N \rrbracket & \rightarrow & \mathbb{C} \, , \\
(y, i) & \mapsto & f(y, i) \, .
\end{array}
\right.
\end{equation}
For each $i \in \llbracket 1, N \rrbracket$, we define a function:
\begin{equation}
f_i: \left\{
\begin{array}{ccc}
[0, \ell] & \rightarrow & \mathbb{C} \, , \\
y & \mapsto & f_i(y) \equiv f(y, i) \, .
\end{array}
\right.
\end{equation}
\begin{itemize}
\item $f$ is continuous/differentiable at $(y,i)=(y_0,i_0)$ if $f_{i_0}$ is continuous/differentiable at $y=y_0$. The derivative of $f$ at $(y,i)=(y_0,i_0)$ is $\partial_y f(y,i) \equiv \partial_y f_i(y)$.
\item $f$ is continuous across the junction if
\begin{equation}
\forall (i,j), \ f(0,i)=f(0,j) \, .
\end{equation}
If it is not the case, $f$ is discontinuous/multivalued at the junction.
\end{itemize}

\paragraph{Test function on $\mathcal{K}_N$ --}
The set of test functions $\mathcal{T}$ is the set of all complex functions $\varphi$ on $\mathcal{K}_N$ such that the functions $\varphi_i$ are infinitely differentiable bounded functions on $[0, \ell]$. We stress that a function $\varphi \in \mathcal{T}$ and/or its derivatives can be discontinuous at the junction.

\paragraph{Distribution --}
A distribution $D \in \mathcal{T}'$ is a linear form on $\mathcal{T}$:
\begin{equation}
\forall \varphi \in \mathcal{T} \, , \ D[\varphi] \equiv \sum_{i=1}^N D_i[\varphi_i] \, ,
\end{equation}
where for every compact set $\mathfrak{B}_i \in [0, \ell]$, there exist constants $C_i$ and $m_i$ such that
\begin{equation}
\forall \varphi \in \mathcal{T} \, , \ \text{supp}(\varphi_i) \in \mathfrak{B}_i \, , \  \left| D_i[\varphi_i] \right| \leq C_i \sum_{\alpha_i\leq m_i} \sup \left| \partial_y^{\alpha_i} \varphi_i (y, i) \right| \, .
\end{equation}

\paragraph{Regular distribution --}
For any integrable complex function  $f$ on $\mathcal{K}_N$, one can define a regular distribution $\widetilde{f} \in \mathcal{T}'$ such that
\begin{equation}
\forall \varphi \in \mathcal{T} \, , \ \widetilde{f}[\varphi] \equiv \sum_{i=1}^N \widetilde{f}_i[\varphi_i] \ \ \ \text{with} \ \ \ \widetilde{f}_i[\varphi_i] \equiv \int_{0}^{\ell} dy \ f(y, i) \, \varphi(y, i) \, .
\end{equation}
A distribution which is not regular is singular.

\paragraph{Product of distributions --}
If $D \in \mathcal{T}'$ and $f \in \mathcal{T}$, one can define the product $f D$ as
\begin{equation}
(fD)[\varphi] \equiv D[f \varphi] \, .
\end{equation}
If $\widetilde{f}$ is the regular distribution associated to $f$, the product $\widetilde{f} D$ is defined as
\begin{equation}
\widetilde{f}D \equiv fD \, .
\end{equation}

\paragraph{Dirac distribution --}
The Dirac distribution on $\mathcal{K}_N$ centered at $(y_0,i_0)$ is the singular distribution $\delta_{y_0,i_0}$ defined as
\begin{equation}
\forall \varphi \in \mathcal{T} \, , \ \delta_{y_0,i_0}[\varphi] \equiv  \varphi(y_0, i_0) \, .
\end{equation}
We want to build a Dirac-like distribution $\delta_{J/V}$ centered at $J/V$ to localize interactions at the junction. Consider the $N$-star $\mathcal{S}_N$. Let $\eta$ be an infinitely differentiable real function on $\mathcal{S}_N$ such that
\begin{equation}
\forall y \in [0, \ell] \, , \ \forall (i, j) \in \llbracket 1, N \rrbracket^2 \, , \ \eta(y,i) = \eta(y,j) \, ,
\label{def_eta}
\end{equation}
and
\begin{equation}
\sum_{i=1}^N \int_0^\ell dy \ \eta(y,i) = 1 \, .
\label{norm_eta}
\end{equation}
We define $\eta_\epsilon$:
\begin{equation}
\eta_\epsilon (y,i) = \dfrac{1}{\epsilon} \, \eta \left( \dfrac{y}{\epsilon}, i \right) \, ,
\label{dilat_eta}
\end{equation}
with $\epsilon > 0$, and we associate the regular distribution $\widetilde{\eta_\epsilon}$ to it. The Dirac distribution $\delta_J$ at the junction $J$ is defined as the weak limit:
\begin{equation}
\delta_J \equiv \lim_{\epsilon \to 0} \eta_\epsilon \, .
\end{equation}
We have
\begin{align}
\forall \varphi \in \mathcal{T} \, , \ \delta_J[\varphi]
&= \lim_{\epsilon \to 0} \sum_{i=1}^N \int_0^\ell dy \ \eta_\epsilon (y,i) \, \varphi (y,i) \, , \nonumber \\
&= \lim_{\epsilon \to 0} \sum_{i=1}^N \int_0^\ell dy \ \eta (y,i) \, \varphi ( \epsilon y,i) \, , \nonumber \\
&= \sum_{i=1}^N \varphi(0,i) \int_0^\ell dy \ \eta(y,i) \, , \nonumber \\
&= \dfrac{1}{N} \, \sum_{i=1}^N \varphi(0,i) \, .
\end{align}
We conclude that
\begin{equation}
\delta_J = \dfrac{1}{N} \, \sum_{i=1}^N \delta_{0,i} \, .
\end{equation}
In the same way, one can build a Dirac distribution $\delta_V$ at the vertex $V$ of the $N$-rose $\mathcal{R}_N$ such that
\begin{equation}
\delta_V \equiv \lim_{\epsilon \to 0} \eta_\epsilon \, ,
\end{equation}
where $\eta_\epsilon$ is defined as in Eq.~\eqref{dilat_eta} and $\eta$ is an infinitely differentiable real function on $\mathcal{R}_N$ such that
\begin{equation}
\forall y \in [0, \ell] \, , \ \forall (i, j) \in \llbracket 1, N \rrbracket^2 \, , \ \eta(y,i)=\eta(y,j) \ \text{and} \ \eta(y-\ell,i)=\eta(y-\ell,j) \, ,
\end{equation}
and normalized as in Eq.~\ref{norm_eta}. Then,
\begin{equation}
\delta_V = \dfrac{1}{2N} \, \sum_{i=1}^N \left( \delta_{0,i} + \delta_{\ell,i} \right) \, .
\end{equation}
We have thus defined a Dirac distribution centered at the junction $J/V$ which acts on test functions, possibly discontinuous at $J/V$.

\paragraph{Distributional derivative --} In Kurasov's distribution theory, one defines a distributional derivative in the same way as in Schwartz's distribution theory. The distributional derivative $\partial_y D$ of a distribution $D \in \mathcal{T}'$ is defined by
\begin{equation}
\forall \varphi \in \mathcal{T} \, , \ \partial_y D [\varphi] = - D [\partial_y \varphi] \, .
\end{equation}
The derivative of a regular distribution $\widetilde{f} \in \mathcal{T}'$ is thus
\begin{equation}
\partial_y \widetilde{f} = \left\{ \partial_y f \right\} + \sum_{i=1}^N \left( \delta_{0, i} - \delta_{\ell, i} \right) f \, ,
\end{equation}
where $\left\{ \partial_y f \right\}$ is the regular distribution associated to derivative $\partial_y f$. As in original Kurasov's distribution theory, the distributional derivative does not coincide with the derivative defined in the classical sense. For instance, the distributional derivative of the regular distribution associated to a constant function is not zero. For the unit function $\mathbf{1}: (y, i) \mapsto 1$, we have
\begin{equation}
\partial_y \widetilde{\mathbf{1}} = \sum_{i=1}^N \left( \delta_{0, i} - \delta_{\ell, i} \right) \, .
\end{equation}
Instead, it would be more natural to define the distributional derivative as
\begin{equation}
\forall \varphi \in \mathcal{T} \, , \ \partial_y D [\varphi] = - D [\partial_y \varphi] - \sum_{i=1}^N \left[ \left( \delta_{0, i} - \delta_{\ell, i} \right) D \right] [\varphi] \, .
\end{equation}
However, in this case, a distributional derivative of the Dirac distributions $\delta_{0/\ell, i}$ and $\delta_{J/V}$ would involve Dirac distributions squared which is not defined. Therefore, the price to pay in order to define a useful distributional derivative is to have extra boundary terms at the vertices, compared to the traditional distributional derivative for a regular distribution in Schwartz's distribution theory. One can thus define the $n^\text{th}$ derivative of the Dirac distribution $\delta_{y_0, i_0}$ as
\begin{equation}
\partial_y^n \delta_{y_0, i_0} [\varphi] =  (-1)^n \, \partial_y^n \varphi(y_0, i_0) \, .
\end{equation}

\paragraph{Moment expansion --}
We will adapt the moment expansion \cite{Estrada:1994} of Schwartz's distribution theory to our case. Consider the $N$-star. The Taylor series of a test function $\varphi \in \mathcal{T}$ is
\begin{equation}
\forall y \in [0, \ell], \forall i \in \llbracket 1, N \rrbracket, \ \varphi(y,i) = \sum_{n=0}^{+\infty} \partial_y^n \varphi(0,i) \, \dfrac{y^n}{n!} \, .
\end{equation}
Then, the action of the previous regular distribution $\widetilde{\eta}$ is
\begin{equation}
\widetilde{\eta} [\varphi] = \sum_{i=1}^N \int_0^\ell dy \ \eta(y,i) \sum_{n=0}^{+\infty} \partial_y^n \varphi(0,i) \, \dfrac{y^n}{n!} \, .
\end{equation}
We define the $n^\text{th}$ moment of the function $\eta$ as
\begin{equation}
\mu_n = \widetilde{\eta} [y^n] = \sum_{i=1}^N \int_0^\ell dy \ \eta(y,i) \, \dfrac{y^n}{n!} \, .
\end{equation}
Thus,
\begin{align}
\widetilde{\eta} [\varphi] &= \left( \sum_{n=0}^{+\infty} \sum_{i=1}^N \dfrac{(-1)^n \, \mu_n}{N n!} \, \partial_y^n \delta_{0,i} \right) [\varphi] \, , \nonumber \\
&= \left( \sum_{n=0}^{+\infty} \dfrac{(-1)^n \, \mu_n}{n!} \, \partial_y^n \delta_J \right) [\varphi] \, .
\end{align}
A similar result is obtained with the $N$-rose. We define the moment expansion of $\widetilde{\eta}$ by
\begin{equation}
\widetilde{\eta} = \sum_{n=0}^{+\infty} \dfrac{(-1)^n \, \mu_n}{n!} \, \partial_y^n \delta_{J/V} \, .
\end{equation}

\section{5D Klein-Gordon Field on a Star/Rose Graph}
\label{KG_field}
The KK-mass spectrum and wave functions of a 5D massless real scalar field on a star/rose graph were studied in Ref.~\cite{Kim:2005aa}. In this section, we generalize it by adding a 5D mass to the scalar field. Besides, we clarify the method and hypotheses of this previous study, especially the hypothesis of continuity of the scalar field across the central vertex of the star/rose graph.

\subsection{Klein-Gordon Equation \& Junction/Boundary Conditions}
\label{KG_field_star}
We study a 5D real scalar field $\Phi$ of mass dimension 3/2 and of mass $M_\Phi$ defined on $\mathcal{M}_4 \times \mathcal{K}_N$. The 5D fields $\Phi_i$ are supposed to be smooth functions on the interval $[0, \ell]$. We associate to $\Phi$ a regular distribution $\widetilde{\Phi}$. The Lagrangian $\widetilde{\mathcal{L}_\Phi}$ describing the dynamics of $\Phi$ is defined at the level of distributions. The action is
\begin{equation}
S_\Phi = \int d^4x \ \widetilde{\mathcal{L}_\Phi} [\mathbf{1}] \, ,
\end{equation}
with the unit test function $\mathbf{1}: (y, i) \mapsto 1$, and
\begin{equation}
\widetilde{\mathcal{L}_\Phi} = - \dfrac{1}{2} \, \widetilde{\Phi} \Box_5 \widetilde{\Phi} - \dfrac{1}{4} \, \widetilde{\Phi}^2 \, \partial_y \left( \delta_{\ell, i} - \delta_{0, i} \right) - \dfrac{M_\Phi^2}{2} \, \widetilde{\Phi}^2 \, ,
\label{L_Phi_star}
\end{equation}
with $M_\Phi^2 \geq 0$. We do not include brane-localized kinetic/mass/interaction terms. The 5D kinetic term is written in a form which is well-defined in distribution theory\footnote{Indeed a 5D kinetic term written as $\dfrac{1}{2} \, \partial^M \widetilde{\Phi} \partial_M \widetilde{\Phi}$ would involve Dirac distributions squared at the vertices, which are not defined.}. We have included specific boundary terms in order to reduce $S_\Phi$ to the standard form for a Klein-Gordon field (with the function $\Phi$):
\begin{equation}
S_\Phi = \int d^4x \ \sum_{i=1}^N \int_0^\ell dy \left( \dfrac{1}{2} \, \partial^M \Phi \partial_M \Phi - \dfrac{M_\Phi^2}{2} \, \Phi^2 \right) \, ,
\label{S_Phi_star}
\end{equation}
where the boundary terms coming from the distributional derivatives cancel each other.\\

We apply Hamilton's principle to the action $S_\Phi$. The procedure is standard (cf. Refs.~\cite{Lalak:2001fd, Csaki:2003dt, Carena:2005gq} for examples with details): the variation of the action in the bulk and on the branes vanish independently with arbitrary field variations; one gets the Euler-Lagrange equations in the bulk and the natural boundary/junction conditions at the brane positions\footnote{A natural boundary/junction condition at a brane position is defined as the condition which cancel the variation of the action on the brane with an arbitrary field variation \cite{Hilbert, Csaki:2003dt, Cheng:2010pt, Angelescu:2019viv, Nortier:2020xms}.}. From the variation of $S_\Phi$ in the bulk, we get the Klein-Gordon equation in each leaf/petal:
\begin{equation}
\left( \partial^M \partial_M + M_\Phi^2 \right) \Phi \left( x^\mu, y, i \right) = 0 \, .
\label{KG_Phi_star}
\end{equation}
For arbitrary $\delta \Phi$'s at the boundaries of the star graph, we obtain Neumann boundary conditions on the $B_i$-branes:
\begin{equation}
\partial_{y} \Phi \left( x^\mu, \ell, i \right) = 0 \, .
\label{BCs_Phi_star}
\end{equation}
The natural junction condition on the $J/V$-brane depends on the continuity hypothesis of the 5D field across this brane. Remember that $\Phi$ can be continuous or discontinuous across the $J/V$-brane. This feature depends on the microscopic structure of the $J/V$-brane in the UV-completion, and is a model building hypothesis in a bottom-up approach. The $\delta \Phi$'s inherite the (dis)continuity properties from $\Phi$ across the $J/V$-brane, and we extract the junction condition for arbitrary $\delta \Phi$'s:
\begin{itemize}
\item If $\Phi$ is allowed to be discontinuous across the junction, we get Neumann junction conditions:
\begin{equation}
\partial_y \Phi(x^\mu, 0/\ell, i) = 0 \, .
\end{equation}
We say that the $J/V$-brane is \textit{airtight} to the field $\Phi$, which means that the spectrum is equivalent to the one obtained by disconnecting the $N$ bonds at the vertex $J/V$ into $N$ disjointed intervals. There is a basis where a KK-mode has its wave function which is normalizable in only one leaf/petal and is vanishing in the others. A brane which is airtight to the field behaves like a boundary for this field.
\item If we impose to $\Phi$ to be continuous across the junction, we get a Neumann-Kirchhoff junction condition:
\begin{equation}
\left\{
\begin{array}{rcl}
\displaystyle{\sum_{i=1}^N \partial_{y} \Phi \left( x^\mu, 0, i \right) = 0} & \text{for} & \mathcal{K}_N = \mathcal{S}_N \, , \\ \\
\displaystyle{\sum_{i=1}^N \left[ \partial_{y} \Phi \left( x^\mu, y, i \right) \right]_{y=0}^{\ell} = 0} & \text{for} & \mathcal{K}_N = \mathcal{R}_N \, ,
\end{array}
\right.
\label{JC_Phi_star}
\end{equation}
with $\left[ g(y) \right]_{y=a}^b = g(b) - g(a)$. As we will see, when $\Phi$ is continuous across the junction, the wave function of a KK-mode is nonvanishing in several leaves/petals so the $J/V$-brane does not behave as a boundary.
\end{itemize}

\subsection{Kaluza-Klein Dimensional Reduction}
\label{KK_scalar_20}
Here we will not study the KK-dimensional reduction when $\Phi$ is allowed to be discontinuous across the junction, since it reduces to a 5D scalar field on $N$ disjointed intervals. The case of a 5D scalar field on an interval is very well known in the literature \cite{Dobrescu:2008zz}. In the following KK-mode analysis, we focus only on the case where $\Phi$ is continuous at the junction, so we have the Neumann-Kirchhoff junction condition \eqref{JC_Phi_star}.

\subsubsection{Separation of Variables}
We perform the KK-dimensional reduction of the 5D field theory to an effective 4D one in terms of KK-degrees of freedom. A general 5D field $\Phi$ can be expanded as
\begin{equation}
\Phi \left( x^\mu, y, i \right) = \sum_b \ \sum_{n_b} \ \sum_{d_b} \phi^{(b, \, n_b, \, d_b)} \left( x^\mu \right) \, f_\phi^{(b, \, n_b, \, d_b)} \left( y, i \right) \, .
\label{KK_Phi_star}
\end{equation}
We label each KK-mode by a triplet $(b, n_b, d_b)$, where:
\begin{itemize}
\item $b$ labels the different KK-towers for the same 5D field, which are defined by different mass spectra (see below);
\item $n_b$ labels the levels in the KK-tower $b$;
\item $d_b$ labels the degenerate modes for each KK-level $(b, n_b)$. We choose the notation $d_b$, instead of the more appropriate one $d(b, n_b)$, for simplifying the notations since we will see that each KK-level for a KK-tower $b$ has the same degeneracy: there is thus no ambiguity.
\end{itemize}

The 5D equation \eqref{KG_Phi_star} splits into the Klein-Gordon equations for the 4D fields $\phi^{(b, \, n_b, \, d_b)}$:
\begin{equation}
\left( \partial^\mu \partial_\mu + \left[ m_\phi^{(b, \, n_b)} \right]^2 \right) \phi^{(b, \, n_b, \, d_b)} (x^\mu) = 0 \, ,
\label{KG_KK-Phi_star}
\end{equation}
with
\begin{equation}
\left[ m_\phi^{(b, \, n_b)} \right]^2 = M_\Phi^2 + \left[ k_\phi^{(b, \, n_b)} \right]^2 \, ,
\label{m_func_k}
\end{equation}
and the differential equations for the wave functions $f_\phi^{(b, \, n_b, \, d_b)}$:
\begin{equation}
\left( \partial_{y}^2 + \left[ k_\phi^{(b, \, n_b)} \right]^2 \right) f_\phi^{(b, \, n_b, \, d_b)} \left( y, i \right) = 0 \, ,
\label{wave_eq_Phi_star}
\end{equation}
where $\left[ m_\phi^{(b, \, n_b)} \right]^2 \geq 0$ is the mass squared of the KK-modes $\phi^{(b, \, n_b, \, d_b)}$, and $\left[ k_\phi^{(b, \, n_b)} \right]^2 \in [0, \ + \infty)$ is an eigenvalue of the operator $\partial_{y}^2$ on $\mathcal{K}_N$ associated to the eigenfunctions $f_\phi^{(b, \, n_b, \, d_b)}$. The orthonormalization conditions for the wave functions $f_\phi^{(b, \, n_b, \, d_b)}$ are
\begin{equation}
\sum_{i=1}^N \int_0^\ell dy \ f_\phi^{(b, \, n_b, \, d_b)}(y, i) \, f_\phi^{(b', \, n'_{b'}, \, d'_{b'})}(y, i) = \delta^{bb'} \, \delta^{n_{b} n'_{b'}} \, \delta^{d_{b} d'_{b'}} \, .
\label{norm_wave_Phi_star}
\end{equation}
The conditions on the 5D field $\Phi$ on the 3-branes are naturally transposed to conditions on the KK-wave functions $f_\phi^{(b, \, n_b, \, d_b)}$.

\subsubsection{Zero-Modes}
We are looking for zero-mode solutions ($b=0$, $n_0 = 0$, $k_\phi^{(0, \, 0)}=0$) of Eq.~\eqref{wave_eq_Phi_star}. For both compactifications on $\mathcal{S}_N$ and $\mathcal{R}_N$, there is only one zero-mode ($d_0 \in \{1\}$) propagating in the whole graph whose wave function is flat, such that:
\begin{equation}
f_\phi^{(0, \, 0, \, 1)} (y, i) = \sqrt{\dfrac{1}{N \ell}} \, .
\label{zero_mode_sol_Phi_2}
\end{equation}

\subsubsection{Excited Modes}

\subsubsection*{\boldmath \textcolor{black}{$a)$ $N$-Star}}
The general solutions of Eq.~\eqref{wave_eq_Phi_star}, satisfying the Neumann boundary conditions \eqref{BCs_Phi_star}, are of the form:
\begin{equation}
f_\phi^{(b, \, n_b, \, d_b)} (y, i) = A_i^{(b, \, n_b, \, d_b)} \, \cos \left[ k_\phi^{(b, \, n_b)} \, (y-\ell) \right] \, ,
\end{equation}
with $A_i^{(b, \, n_b, \, d_b)} \in \mathbb{R}$. The continuity condition on the wave functions at the $J$-brane gives
\begin{equation}
\forall (i,j) \, , \ A_i^{(b, \, n_b, \, d_b)} \, \cos \left[ k_\phi^{(b, \, n_b)} \, \ell \right] = A_j^{(b, \, n_b, \, d_b)} \, \cos \left[ k_\phi^{(b, \, n_b)} \, \ell \right] \, ,
\label{cont_phi_star_20}
\end{equation}
which leads to two kinds of excited KK-modes: the KK-wave functions $f_\phi^{(b, \, n_b, \, d_b)}$ can vanish or not on the $J$-brane.

\subsubsection*{\boldmath \textcolor{black}{\textit{First case: $f_\phi^{(b, \, n_b, \, d_b)}(0,i)=0$}}}
Eq.~\eqref{cont_phi_star_20} gives the KK-mass spectrum
\begin{equation}
\cos \left[ k_\phi^{(b, \, n_b)} \, \ell \right] = 0 \ \ \ \underset{b=1}{\Longrightarrow} \ \ \ k_\phi^{(1, \, n_1)} = \left( n_1 + \dfrac{1}{2} \right) \dfrac{\pi}{\ell} \, , \ n_1 \in \mathbb{N} \, ,
\label{mass_spect_Phi_star_2}
\end{equation}
which defines the KK-tower $b=1$. The Neumann-Kirchhoff junction condition \eqref{JC_Phi_star} implies
\begin{equation}
\sum_{i=1}^N A_i^{(1, \, n_1, \, d_1)} = 0 \, .
\label{eq_Ai_20}
\end{equation}
Each KK-level $(1, n_1)$ is thus $N-1$ times degenerate ($d_1 \in \llbracket 1, N-1 \rrbracket$) and the KK-wave functions are
\begin{equation}
f_\phi^{(1, \, n_1, \, d_1)} (y, i) = \epsilon^{(d_1)}_i \sqrt{\dfrac{2}{\ell}} \, \cos \left[ k_\phi^{(1, \, n_1)} \, (y-\ell) \right] \, ,
\end{equation}
with the $(N-1)$-vector basis:
\begin{align}
\overrightarrow{\epsilon^{(1)}} &= \dfrac{1}{\sqrt{2}} \left( 1, -1, 0, \cdots, 0 \right) \, , \nonumber \\
\overrightarrow{\epsilon^{(2)}} &= \dfrac{1}{\sqrt{6}} \left( 1, 1, -2, 0, \cdots, 0 \right) \, , \nonumber \\
\vdots \nonumber \\
\overrightarrow{\epsilon^{(N-1)}} &= \dfrac{1}{\sqrt{N(N-1)}} \left( 1, 1, \cdots, 1, -(N-1) \right) \, .
\label{basis_epsi}
\end{align}
In this basis, most of these modes do not propagate in all leaves.

\subsubsection*{\boldmath \textcolor{black}{\textit{Second case: $f_\phi^{(b, \, n_b, \, d_b)}(0,i) \neq 0$}}}
We have $\cos \left[ k_\phi^{(b, \, n_b)} \, \ell \right] \neq 0$ so Eq.~\eqref{cont_phi_star_20} gives
\begin{equation}
\forall (i, j) \, , \ A_i^{(b, \, n_b, \, d_b)} = A_j^{(b, \, n_b, \, d_b)} \equiv A^{(b, \, n_b, \, d_b)} \, .
\label{eq_A_20}
\end{equation}
The Kirchhoff junction condition \eqref{JC_Phi_star} leads to the KK-mass spectrum
\begin{equation}
\sin \left[ k_\phi^{(b, \, n_b)} \, \ell \right] = 0 \ \ \ \underset{b=2}{\Longrightarrow} \ \ \ k_\phi^{(2, \, n_2)} = n_2 \, \dfrac{\pi}{\ell} \, , \ n_2 \in \mathbb{N}^* \, ,
\label{mass_spect_Phi_star_1}
\end{equation}
which defines the KK-tower with $b=2$ whose KK-levels are not degenerate ($d_2 \in \{1\}$). The KK-wave functions are
\begin{equation}
f_\phi^{(2, \, n_2, \, 1)} (y, i) = \sqrt{\dfrac{2}{N\ell}} \, \cos \left[ k_\phi^{(2, \, n_2)} \, (y-\ell) \right] \, ,
\end{equation}
thus these modes propagate in the whole star graph.

\subsubsection*{\boldmath \textcolor{black}{$b)$ $N$-Rose}}
Again, to satisfy the continuity condition at the $V$-brane, the KK-wave functions $f_\phi^{(b, \, n_b, \, d_b)}$ can vanish or not at the vertex.

\subsubsection*{\boldmath \textcolor{black}{\textit{First case: $f_\phi^{(b, \, n_b, \, d_b)}(0,i)=0$}}}
The general solutions of Eq.~\eqref{wave_eq_Phi_star} with $f_\phi^{(b, \, n_b, \, d_b)}(0,i)=0$ are of the form:
\begin{equation}
f_\phi^{(b, \, n_b, \, d_b)} (y, i) = A_i^{(b, \, n_b, \, d_b)} \, \sin \left[ k_\phi^{(b, \, n_b)} \, y \right] \, ,
\label{gen_prof_rose_50}
\end{equation}
with $A_i^{(b, \, n_b, \, d_b)} \in \mathbb{R}$. The periodicity condition for each petal at the vertex gives $\sin \left[ k_\phi^{(b, \, n_b)} \, \ell \right] = 0$. Moreover, the Neumann-Kirchhoff junction condition \eqref{JC_Phi_star} implies
\begin{equation}
\left( \cos \left[ k_\phi^{(b, \, n_b)} \, \ell \right] - 1 \right) \sum_{i=1}^N A_i^{(b, \, n_b, \, d_b)} = 0 \, .
\label{JC_phi_30}
\end{equation}
\newpage

There are thus two possibilities:
\begin{itemize}
\item First possibility:\\
The KK-mass spectrum is
\begin{equation}
\left\{
\begin{array}{rcl}
\cos \left[ k_\phi^{(b, \, n_b)} \, \ell \right] &\neq& 1 \\
\sin \left[ k_\phi^{(b, \, n_b)} \, \ell \right] &=& 0
\end{array}
\right.
\ \ \ \underset{b=1}{\Longrightarrow} \ \ \ k_\phi^{(1, \, n_1)} = (2n_1+1) \, \dfrac{\pi}{\ell} \, , \ n_1 \in \mathbb{N} \, ,
\label{mass_spect_Phi_rose_3}
\end{equation}
and defines the KK-tower $b=1$. From the conditions \eqref{JC_phi_30}, we get Eq.~\eqref{eq_Ai_20} so $d_1 \in \llbracket 1, N-1 \rrbracket$ and the KK-wave functions are
\begin{equation}
f_\phi^{(1, \, n_1, \, d_1)} (y, i) = \epsilon^{(d_1)}_i \sqrt{\dfrac{2}{\ell}} \, \sin \left[ k_\phi^{(1, \, n_1)} \, y \right] \, ,
\end{equation}
with the $(N-1)$-vector basis \eqref{basis_epsi} for which most of the modes do not propagate in all petals.
\item Second possibility:\\
The KK-mass spectrum is
\begin{equation}
\left\{
\begin{array}{rcl}
\cos \left[ k_\phi^{(b, \, n_b)} \, \ell \right] &=& 1 \\
\sin \left[ k_\phi^{(b, \, n_b)} \, \ell \right] &=& 0
\end{array}
\right.
 \ \ \ \underset{b=2}{\Longrightarrow} \ \ \ k_\phi^{(2, \, n_2)} = 2n_2 \, \dfrac{\pi}{\ell} \, , \ n_2 \in \mathbb{N}^* \, ,
\label{mass_spect_Phi_rose_2}
\end{equation}
which satisfies Eq.~\eqref{JC_phi_30} and defines the KK-tower $b=2$. We get $d_2 \in \llbracket 1, N \rrbracket$ and the KK-wave functions are
\begin{equation}
f_\phi^{(2, \, n_2, \, d_2)} (y, i) = \eta^{(d_2)}_i \sqrt{\dfrac{2}{\ell}} \, \sin \left[ k_\phi^{(2, \, n_2)} \, y \right] \, ,
\label{wave_funct_1}
\end{equation}
with the $N$-vector basis:
\begin{align}
\overrightarrow{\eta^{(1)}} &= \left( 1, 0, \cdots, 0 \right) \, , \nonumber \\
\overrightarrow{\eta^{(2)}} &= \left( 0, 1, 0, \cdots, 0 \right) \, , \nonumber \\
\vdots \nonumber \\
\overrightarrow{\eta^{(N)}} &= \left( 0, \cdots, 0, 1 \right) \, .
\label{basis_eta}
\end{align}
\end{itemize}
Each mode propagates in only one petal.

\subsubsection*{\boldmath \textcolor{black}{\textit{Second case}: $f_\phi^{(b, \, n_b, \, d_b)}(0,i) \neq 0$}}
The KK-mass spectrum is the same as in Eq.~\eqref{mass_spect_Phi_rose_2}. Each KK-level $(b, n_b)$ is thus degenerate with the $N$ KK-modes of the level $(2, n_2)$ so $d_2 \in \llbracket 1, N+1 \rrbracket$: we label the KK-modes with nonvanishing wave functions at the junction by the triplet $(2, n_2, N+1)$. The KK-wave functions are
\begin{equation}
f_\phi^{(2, \, n_2, \, N+1)} (y, i) = \sqrt{\dfrac{2}{N \ell}} \, \cos \left( m_\phi^{(2, \, n_2)} \, y \right) \, .
\label{wave_phi_rose_interm}
\end{equation}
These modes propagate in the whole rose graph.\\

Finally, we insist on the fact that all KK-towers labeled by $b$ are present in the spectrum: they do not correspond to different models. The 5D field $\Phi$ has one zero-mode of mass $M_\Phi$ in both geometries ($\mathcal{K}_N = \mathcal{S}_N$ or $\mathcal{R}_N$) and excited modes. For a massless 5D field ($M_\Phi=0$), the mass gap between the KK-modes is of the order of $1/\ell$. Some KK-modes have wave functions which vanish at the junction. We will see a physical application of these results in Subsection~\ref{pheno_star_rose_graviton}, where we will study a toy model of a 5D spinless graviton which is just a real scalar field with $M_\Phi = 0$. The zero-mode is thus identified with the 4D massless graviton (where we do not take into account the spin).

\section{5D Massless Dirac Field  on a Star/Rose Graph}
\label{Dirac_field}
Recently, a 5D Dirac field with a compactification on a flat rose graph was considered in Ref.~\cite{Fujimoto:2019fzb}. They took petals of possibly different circumferences and included a 5D Dirac mass for the fermion. In this framework, they considered the rose graph as a master quantum graph, since one can reduce it to a star graph by a suitable choice of junction conditions. They studied the general mathematical properties of the junction conditions for the rose graph and classified them. Their work was restricted to the analysis of the zero-modes only: the KK-mass spectrum and wave functions of the excited modes were not considered. They determined the number of zero-mode solutions for each type of boundary conditions in their classification. Their work was motivated by the future goals of generating three fermion generations and the features of the flavor sector of the SM-fermions from the zero-modes of only one generation of 5D fermions. In this section, we study the particular case of a 5D massless Dirac field on a star/rose graph with identical leaves/petals. We use a different approach compared to the one of Ref.~\cite{Fujimoto:2019fzb}. Instead of imposing arbitrary junction conditions, we keep only the natural junction conditions at the vertices, i.e. the junction conditions for which the variation of the action at the vertices vanishes for arbitrary field variations \cite{Hilbert, Csaki:2003dt, Cheng:2010pt, Angelescu:2019viv, Nortier:2020xms}. Indeed, we prefer junction conditions originating from the variation of the action (and thus of the fields) at the vertices. We will see that the natural junction conditions depend only on the hypothesis of the continuity of the fields at the junction. In this approach, we need the Henningson-Sfetsos (HS) boundary terms for 5D fermions \cite{Henningson:1998cd, Mueck:1998iz, Arutyunov:1998ve, Henneaux:1998ch, Contino:2004vy, vonGersdorff:2004eq, vonGersdorff:2004cg, Angelescu:2019viv, Nortier:2020xms} whose importance was stressed recently in Refs.~\cite{Angelescu:2019viv, Nortier:2020xms}. Besides, we do not restrict ourselves to the study of the zero-modes only; we determine the KK-mass spectrum and wave functions of all KK-modes.

\subsection{Dirac-Weyl Equations \& Junction/Boundary Conditions}
We study a 5D massless Dirac field
\begin{equation}
\Psi =
\begin{pmatrix}
\Psi_L \\
\Psi_R
\end{pmatrix}
\end{equation}
of mass dimension 2 defined on $\mathcal{M}_4 \times \mathcal{K}_N$, where the fields $\Psi_L$ and $\Psi_R$ describe fermion fields of left and right-handed 4D chirality respectively. To the function $\Psi$, we associate the regular distribution $\widetilde{\Psi}$. The action is
\begin{equation}
S_\Psi = \int d^4x \ \widetilde{\mathcal{L}_\Psi} [\mathbf{1}] \, ,
\end{equation}
with the Lagrangian
\begin{equation}
\widetilde{\mathcal{L}_\Psi}
= \dfrac{i}{2} \, \bar{\widetilde{\Psi}} \Gamma^M \overleftrightarrow{\partial_M} \widetilde{\Psi} + \sum_{i=1}^N \dfrac{s_i}{2} \, \bar{\widetilde{\Psi}} \widetilde{\Psi} \, \left( \delta_{0,i} - \delta_{\ell,i} \right) \, ,
\end{equation}
where $\bar{\widetilde{\Psi}} = \Gamma^0 \widetilde{\Psi}$, $\overleftrightarrow{\partial_M} = \vec{\partial_M} - \overleftarrow{\partial_M}$ and $\Gamma^M = \left( \gamma^\mu, i \gamma^5 \right)$ are the 5D Dirac matrices\footnote{Our conventions for the Dirac algebra is given in Appendix~\ref{conventions}.}.
We include the HS boundary terms at the vertices with $s_i=\pm 1$. The relative sign between the HS terms at $(y=0,i)$ and $(y=\ell,i)$ is chosen in order to allow for the existence of zero-modes (cf. Refs.~\cite{Angelescu:2019viv, Nortier:2020xms} for a detailed discussion with a compactification on an interval). If we flip the sign of $s_i$, we exchange the features of the left and right-handed KK-modes. In what follows, we choose $s_i = 1$.\\

The action can be written as
\begin{equation}
S_\Psi = \int d^4x \sum_{i=1}^N \left\{ \left( \int_{0}^{\ell} dy \ \dfrac{i}{2} \, \bar{\Psi} \Gamma^M \overleftrightarrow{\partial_M} \Psi \right) - \left[ \dfrac{1}{2} \, \bar{\Psi} \Psi \right]_{y=0}^{\ell} \right\} \, ,
\label{S_Psi_star}
\end{equation}
where the boundary terms coming from the distributional derivatives cancel each other. The conserved Noether current associated to the symmetry $U(1): \ \Psi \ \mapsto \ e^{-i \alpha} \Psi$, with $\alpha \in \mathbb{R}$, is
\begin{equation}
j_\Psi^M = \bar{\Psi} \Gamma^M \Psi \ \ \ \text{with} \ \ \ \partial_M j_\Psi^M = 0 \, .
\end{equation}
Current conservation requires a Kirchhoff condition for the current at the junction:
\begin{equation}
\left\{
\begin{array}{rcl}
\displaystyle{\sum_{i=1}^N j_\Psi^M \left( x^\mu, 0, i \right) \overset{!}{=} 0} & \text{for} & \mathcal{K}_N = \mathcal{S}_N \, , \\ \\
\displaystyle{\sum_{i=1}^N \left[ j_\Psi^M \left( x^\mu, y, i \right) \right]_{y=0}^{\ell} \overset{!}{=} 0} & \text{for} & \mathcal{K}_N = \mathcal{R}_N \, .
\end{array}
\right.
\end{equation}
For the component $M = 4$ one gets at the junction:
\begin{equation}
\left\{
\begin{array}{rcl}
\displaystyle{\sum_{i=1}^N \left. \left( \Psi_L^\dagger \Psi_R - \Psi_R^\dagger \Psi_L \right) \right|_{y=0} \overset{!}{=} 0} & \text{for} & \mathcal{K}_N = \mathcal{S}_N \, , \\ \\
\displaystyle{\sum_{i=1}^N \left[ \Psi_L^\dagger \Psi_R - \Psi_R^\dagger \Psi_L \right]_{y=0}^{\ell} \overset{!}{=} 0} & \text{for} & \mathcal{K}_N = \mathcal{R}_N \, .
\end{array}
\right.
\label{Kir_current_psi}
\end{equation}

We apply Hamilton's principle to the action $S_\Psi$, with arbitrary variations of the fields $\delta \Psi_{L/R}$ in the bulk and on the branes. We get the massless Dirac-Weyl equations for the 5D fields $\Psi_{L/R}$:
\begin{equation}
\left \{
\begin{array}{r c l}
i \sigma^\mu \partial_\mu \Psi_R  (x^\mu, y, i) + \partial_y \Psi_L  (x^\mu, y, i) &=& 0 \, ,
\\ \vspace{-0.2cm} \\
i \bar{\sigma}^\mu \partial_\mu \Psi_L  (x^\mu, y, i) - \partial_y \Psi_R  (x^\mu, y, i) &=& 0 \, .
\end{array}
\right.
\label{Dirac_Psi_star}
\end{equation}
Therefore, when the fields are on-shell, $\Psi_L$ and $\Psi_R$ are not independent so the junction/boundary conditions must not overconstrain $\Psi_L$ and $\Psi_R$ at the same point (cf. Ref.~\cite{Nortier:2020xms} for a detailed discussion and Refs~\cite{Henneaux:1998ch, Contino:2004vy} for a discussion in a holographic approach). The addition of the HS terms guarantee that only $\Psi_L$ is constrained on the branes by the minimization of the action \cite{Henneaux:1998ch, Contino:2004vy, Angelescu:2019viv, Nortier:2020xms}. For arbitrary field variations at the boundaries of the star graph, we get Dirichlet boundary conditions at the $B_i$-branes:
\begin{equation}
\Psi_L \left( x^\mu, \ell, i \right) = 0 \ \text{(for $\mathcal{K}_N = \mathcal{S}_N)$} \, .
\label{D_psi_B}
\end{equation}
The fields $\Psi_L$ and $\Psi_R$ can be independently (dis)continuous across the junction, which must be stated in the model definition in a bottom-up approach. $\delta \Psi_{L/R}$ is (dis)continuous as $\Psi_{L/R}$. One can explore the different possibilities of junction conditions for $\Psi_L$ with arbitrary field variations at $(y=0,i)$ summarized in Tab.~\ref{table_junction}, depending on the (dis)continuity of the fields here:
\begin{itemize}
\item Case 1: If both $\Psi_L$ and $\Psi_R$ are allowed to be discontinuous, one gets Dirichlet junction conditions:
\begin{equation}
\left\{
\begin{array}{l}
\Psi_L(x^\mu, 0, i) = 0 \ \text{for $\mathcal{K}_N = \mathcal{S}_N$ or $\mathcal{R}_N$,} \\
\Psi_L(x^\mu, \ell, i) = 0 \ \text{for $\mathcal{K}_N = \mathcal{R}_N$,}
\end{array}
\right.
\label{D_junction_psi}
\end{equation}
which correspond to a 5D field $\Psi$ defined on $N$ disjointed intervals. The spectrum of a 5D fermion on an interval is well known in the literature \cite{Csaki:2003sh}. Each KK-mode has a normalizable wave function in only one leaf/petal and is vanishing in the others. The $J/V$-brane is airtight, which is illustrated by the fact that each incoming or outcoming current at the $J/V$-brane vanishes: the junction behaves like a boundary \eqref{Kir_current_psi}. There is a chiral zero-mode (here right-handed) in each leaf/petal and a KK-tower of vector-like fermions with a mass gap of $\pi/\ell$. If one generation of the SM-fermion sector propagates on a $3$-star/$3$-rose it is possible to generate three generations at the level of zero-modes with a airtight $J/V$-brane. The mechanism is the same as in Refs.~\cite{Fujimoto:2012wv, Fujimoto:2013ki, Fujimoto:2014fka, Fujimoto:2014pra, Fujimoto:2017lln, Fujimoto:2019lbo} with point interactions along an interval/circle to generate several zero-modes from a unique discontinuous 5D fermion field.

\item Case 2: If we impose that both $\Psi_L$ and $\Psi_R$ are continuous, the variation of the action at the $V$-brane (rose graph) vanishes without additional condition. However, we get a Dirichlet condition \eqref{D_junction_psi} for $\Psi_L$ at the $J$-brane (star graph).

\item Case 3: If we impose only the continuity on $\Psi_L$, there is a Dirichlet condition \eqref{D_junction_psi} for $\Psi_L$ at the $J/V$-brane.

\item Case 4: If we impose only the continuity on $\Psi_R$, Hamilton's principle gives a Kirchhoff junction condition for $\Psi_L$ at the $J/V$-brane:
\begin{equation}
\left\{
\begin{array}{rcl}
\displaystyle{\sum_{i=1}^N \Psi_L \left( x^\mu, 0, i \right) = 0} & \text{for} & \mathcal{K}_N = \mathcal{S}_N \, , \\ \\
\displaystyle{\sum_{i=1}^N \left[ \Psi_L \left( x^\mu, y, i \right) \right]_{y=0}^{\ell} = 0} & \text{for} & \mathcal{K}_N = \mathcal{R}_N \, .
\end{array}
\right.
\label{JC_Psi_star}
\end{equation}
\end{itemize}
All these cases satisfy the condition \eqref{Kir_current_psi} on the currents.

\begin{table}[h]
\begin{center}
\begin{tabular}{l||c|c}
$N$-star & $\Psi_L$ continuous & $\Psi_L$ discontinuous  \\
\hline \hline
$\Psi_R$ continuous & Dirichlet & Kirchhoff   \\ 
$\Psi_R$ discontinuous & Dirichlet & Dirichlet
\end{tabular}

\vspace{0.5cm}

\begin{tabular}{l||c|c}
$N$-rose & $\Psi_L$ continuous & $\Psi_L$ discontinuous  \\
\hline \hline
$\Psi_R$ continuous &   & Kirchhoff   \\ 
$\Psi_R$ discontinuous & Dirichlet & Dirichlet
\end{tabular}
\caption{The different possibilities of junction conditions for $\Psi_L$ obtained with the action principle and depending on the continuity of $\Psi_L$ and $\Psi_R$. There is no junction condition for $\Psi_R$ from Hamilton's principle because of the HS terms.}
\label{table_junction}
\end{center}
\end{table}

\subsection{Kaluza-Klein Dimensional Reduction}
\label{KK_Drac_20}
For the physical application in Section~\ref{Dirac_Neutrinos}, we will add brane-localized terms at the junction for $\Psi_R$, which are incompatible with the continuity of $\Psi_L$ \cite{Csaki:2003sh, Csaki:2005vy, Nortier:2020xms}. Moreover, we will not consider airtight branes. Therefore, for the following KK-mode analysis, we choose that $\Psi_R$ is continuous at the junction, and $\Psi_L$ is allowed to be discontinuous. We have thus the Kirchhoff junction condition \eqref{JC_Psi_star} for $\Psi_L$.

\subsubsection{Separation of Variables}
In order to perform the KK-dimensional reduction of the 5D field theory, we use the same method as in Subsection~\ref{KK_scalar_20} for the scalar field, with the same system of labels for the KK-modes. We expand the 5D fields $\Psi_{L/R}$ as
\begin{equation}
\left\{
\begin{array}{rcl}
\Psi_{L} \left( x^\mu, y, i \right) &=& \displaystyle{\sum_b \ \sum_{n_b} \ \sum_{d_b} \psi_L^{(b, \, n_b, \, d_b)} \left( x^\mu \right) \, f_L^{(b, \, n_b, \, d_b)} \left( y, i \right)} \, ,  \\
\Psi_{R} \left( x^\mu, y, i \right) &=& \displaystyle{\sum_b \ \sum_{n_b} \ \sum_{d_b} \psi_R^{(b, \, n_b, \, d_b)} \left( x^\mu \right) \, f_R^{(b, \, n_b, \, d_b)} \left( y, i \right)} \, ,
\end{array}
\right.
\label{KK_Psi_star_1}
\end{equation}
where $\psi_{L/R}^{(b, \, n_b, \, d_b)}$ are 4D Weyl fields and $f_{L/R}^{(b, \, n_b, \, d_b)}$ are wave functions defined on $\mathcal{K}_N$. The 5D equations \eqref{Dirac_Psi_star} split into Dirac-Weyl equations for the 4D fields $\psi_{L/R}^{(b, \, n_b, \, d_b)}$:
\begin{equation}
\left \{
\begin{array}{r c l}
i \sigma^\mu \partial_\mu \psi_R^{(b, \, n_b, \, d_b)} (x^\mu) - m_\psi^{(b, \, n_b)} \, \psi_L^{(b, \, n_b, \, d_b)} (x^\mu) &=& 0 \, ,
\\ \vspace{-0.2cm} \\
i \bar{\sigma}^\mu \partial_\mu \psi_L^{(b, \, n_b, \, d_b)} (x^\mu) - m_\psi^{(b, \, n_b)} \, \psi_R^{(b, \, n_b, \, d_b)} (x^\mu) &=& 0 \, ,
\end{array}
\right.
\label{Dirac_KK-Psi_star}
\end{equation}
and the differential equation for the wave functions $f_{L/R}^{(b, \, n_b, \, d_b)}$:
\begin{equation}
\forall y \neq 0 \, , \ \left \{
\begin{array}{r c l}
\partial_y f_R^{(b, \, n_b, \, d_b)} (y, i) - m_\psi^{(b, \, n_b)} \, f_L^{(b, \, n_b, \, d_b)} (y, i) &=& 0 \, ,
\\ \vspace{-0.2cm} \\
\partial_y f_L^{(b, \, n_b, \, d_b)} (y, i) + m_\psi^{(b, \, n_b)} \, f_R^{(b, \, n_b, \, d_b)} (y, i) &=& 0 \, ,
\end{array}
\right.
\label{wave_eq_Psi_star}
\end{equation}
where $m_\psi^{(b, \, n_b)}$ is the mass of the KK-modes $(b, \, n_b, \, d_b)$. The wave functions $f_{L/R}^{(b, \, n_b, \, d_b)}$ are orthonormalized with the conditions
\begin{equation}
\sum_{i=1}^N \int_0^\ell dy \ \left[ f_{L/R}^{(b, \, n_b, \, d_b)}(y, i) \right]^* \, f_{L/R}^{(b', \, n'_{b'}, \, d'_{b'})}(y, i) = \delta^{bb'} \, \delta^{n_{b} n'_{b'}} \, \delta^{d_{b} d'_{b'}} \, ,
\label{orthonorm_KK-Phi_star}
\end{equation}
The conditions on the 5D field $\Psi_{L/R}$ at the vertices are naturally transposed to conditions on the KK-wave functions $f_{L/R}^{(b, \, n_b, \, d_b)}$.

\subsubsection{Zero-Modes}
We are looking for zero-mode solutions ($b=0$, $n_0=0$, $m_\psi^{(0, \, 0)}=0$) of Eq.~\eqref{wave_eq_Psi_star} for which the first order differential equations are decoupled. For both compactifications on $\mathcal{S}_N$ and $\mathcal{R}_N$, there is only one right-handed zero-mode ($d_0 \in \{1\}$) which propagates in the whole graph. Its wave function is continuous across the $J/V$-brane and flat:
\begin{equation}
f_R^{(0, \, 0, \, 1)} (y, i) = \sqrt{\dfrac{1}{N \ell}} \, .
\label{zero_mode_star_psi_20}
\end{equation}
For the left-handed zero-modes, it is necessary to distinguish between the compactification on $\mathcal{S}_N$ and $\mathcal{R}_N$.

\subsubsection*{\boldmath \textcolor{black}{$a)$ $N$-Star}}
There is no left-handed zero-mode for $\mathcal{K}_N = \mathcal{S}_N$.
The theory is thus chiral at the level of the zero-mode, which generalizes the well known result of the particular case of a compactification on the interval $\mathcal{S}_1$. The compactification on a star graph is thus very interesting, since it allows building models where the SM-fields propagate in the EDS: the SM-particles are identified with the zero-modes of the 5D fields. In Section~\ref{Dirac_Neutrinos}, we will identify the right-handed neutrinos with the zero-modes of 5D Dirac fields coupled to brane-localized 4D left-handed neutrinos. The goal is to propose a toy model to obtain small Dirac neutrino masses.

\subsubsection*{\boldmath \textcolor{black}{$b)$ $N$-Rose}}
For $\mathcal{K}_N = \mathcal{R}_N$, we have $N$ degenerate left-handed zero-modes ($d_0 \in \llbracket 1, N \rrbracket$). The theory is vector-like at the level of the zero-modes, which generalizes the result of the compactification on a circle $\mathcal{R}_1$ in the literature. Therefore, with the compactification on a rose graph and without an airtight $V$-brane, one cannot build models with the SM-fields propagating in the EDS except if one is able to suggest a mechanism which generates chirality by giving a mass to the mirror partners of the SM-fermions. If this is possible, one recovers the three SM-generations by taking $N=3$. The KK-wave functions $f_L^{(0, \, 0, \, d_0)}$ are flat in each petal and discontinuous across the $V$-brane (except for $\mathcal{K}_N = \mathcal{R}_1$, the circle, where they can be taken continuous):
\begin{equation}
f_L^{(0, \, 0, \, d_0)} (y, i) = \eta_i^{(d_0)} \sqrt{\dfrac{1}{\ell}} \, ,
\label{zero_mode_f_L}
\end{equation}
with the $N$-vector basis \eqref{basis_eta}: each mode propagates in only one petal.

\subsubsection{Excited Modes}
\label{excited_modes_fermion}
We are looking for massive KK-modes ($m_\psi^{(b, \, n_b)} \neq 0$). The coupled first order differential equations \eqref{wave_eq_Psi_star} can be decoupled into second order ones:
\begin{equation}
\left( \partial_{y}^2 + \left[m_\psi^{(b, \, n_b)}\right]^2 \right) f_{L/R}^{(b, \, n_b, \, d_b)} \left( y, i \right) = 0 \, .
\label{eq_Psi_wave_2nd}
\end{equation}
The KK-wave functions $f_{R}^{(b, \, n_b, \, d_b)}$ are continuous across the junction. In the same way as in the case of the scalar field, it is necessary to distinguish between the cases where the $f_{R}^{(b, \, n_b, \, d_b)}$'s vanish or not at the junction. One can follow the same method as in Subsection~\ref{KK_scalar_20}. We will not give again all the details here since there is no major technical difference. We summarize the results in what follows.

\subsubsection*{\boldmath \textcolor{black}{$a)$ $N$-Star}}
\subsubsection*{\boldmath \textcolor{black}{\textit{First case: $f_R^{(b, \, n_b, \, d_b)}(0,i)=0$}}}
\label{1st_case_psi_star}
The KK-mass spectrum is
\begin{equation}
m_\psi^{(1, \, n_1)} = \left( n_1 + \dfrac{1}{2} \right) \dfrac{\pi}{\ell} \, , \ n_1 \in \mathbb{N} \, ,
\label{mass_spect_Psi_star_2}
\end{equation}
and defines the KK-tower $b=1$. Each KK-level is $N-1$ times degenerate ($d_1 \in \llbracket 1, N-1 \rrbracket$) and the KK-wave functions are
\begin{equation}
\left\{
\begin{array}{rcl}
f_L^{(1, \, n_1, \, d_1)} (y, i) & = & - \epsilon^{(d_1)}_i \sqrt{\dfrac{2}{\ell}} \, \sin \left[ m_\psi^{(1, \, n_1)} \, (y-\ell) \right] \, , \\
f_R^{(1, \, n_1, \, d_1)} (y, i) & = & \epsilon^{(d_1)}_i \sqrt{\dfrac{2}{\ell}} \, \cos \left[ m_\psi^{(1, \, n_1)} \, (y-\ell) \right] \, ,
\end{array}
\right.
\label{Psi_free_fR=0}
\end{equation}
with the $(N-1)$-vector basis \eqref{basis_epsi}. The $f_L^{(1, \, n_1, \, d_1)}$'s are discontinuous across the $J$-brane (except for $\mathcal{K}_N = \mathcal{S}_1$, the interval, where they are taken continuous) and most of them do not propagate in all leaves.

\subsubsection*{\boldmath \textcolor{black}{\textit{Second case: $f_R^{(b, \, n_b, \, d_b)}(0,i) \neq 0$}}}
The KK-mass spectrum is
\begin{equation}
m_\psi^{(2, \, n_2)} = n_2 \, \dfrac{\pi}{\ell} \, , \ n_2 \in \mathbb{N}^* \, ,
\label{mass_spect_Psi_star_1}
\end{equation}
which is not degenerate ($d_2 \in \{1\}$) and defines the KK-tower $b=2$. The KK-wave functions are
\begin{equation}
\left\{
\begin{array}{rcl}
f_L^{(2, \, n_2, \, d_2)} (y, i) &=& - \sqrt{\dfrac{2}{N \ell}} \, \sin \left[ m_\psi^{(2, \, n_2)} \, (y-\ell) \right] \, , \\
f_R^{(2, \, n_2, \, d_2)} (y, i) &=& \sqrt{\dfrac{2}{N \ell}} \, \cos \left[ m_\psi^{(2, \, n_2)} \, (y-\ell) \right] \, ,
\end{array}
\right.
\end{equation}
where the $f_L^{(2, \, n_2, \, d_2)}$'s can be taken continuous across the $J$-brane and propagate in all petals.

\subsubsection*{\boldmath \textcolor{black}{$b)$ $N$-Rose}}
\subsubsection*{\boldmath \textcolor{black}{\textit{First case: $f_R^{(b, \, n_b, \, d_b)}(0,i)=0$}}}
\label{1st_case_psi_rose}
There are two cases:
\begin{itemize}
\item First case: \\
The KK-mass spectrum is
\begin{equation}
m_\psi^{(1, \, n_1)} = (2n_1+1) \, \dfrac{\pi}{\ell} \, , \ n_1 \in \mathbb{N} \, ,
\label{mass_spect_Psi_rose_3}
\end{equation}
and defines the KK-tower $b=1$ with $d_1 \in \llbracket 1, N-1 \rrbracket$. The KK-wave functions are
\begin{align}
\left\{
\begin{array}{rcl}
f_L^{(1, \, n_1, \, d_1)} (y, i) &=& \epsilon^{(d_1)}_i \sqrt{\dfrac{2}{\ell}} \, \cos \left[ m_\psi^{(1, \, n_1)} \, y \right] \, , \\
f_R^{(1, \, n_1, \, d_1)} (y, i) &=& \epsilon^{(d_1)}_i \sqrt{\dfrac{2}{\ell}} \, \sin \left[ m_\psi^{(1, \, n_1)} \, y \right] \, ,
\end{array}
\right.
\label{wave_funct_Psi_rose_fR(0)=0}
\end{align}
with the $(N-1)$-vector basis \eqref{basis_epsi} so most of the modes do not propagate in the whole rose graph. The $f_L^{(1, \, n_1, \, d_1)}$'s are discontinuous across the $V$-brane.
\item Second possibility:\\
The KK-mass spectrum is
\begin{equation}
m_\psi^{(2, \, n_2)} = 2n_2 \, \dfrac{\pi}{\ell} \, , \ n_2 \in \mathbb{N}^* \, ,
\label{mass_spect_Psi_rose_bis}
\end{equation}
and defines the KK-tower $b=2$ with $d_2 \in \llbracket 1, N \rrbracket$. The KK-wave functions are
\begin{equation}
\left\{
\begin{array}{rcl}
f_L^{(2, \, n_2, \, d_2)} (y, i) &=& \eta^{(d_2)}_i \sqrt{\dfrac{2}{\ell}} \, \cos \left[ m_\psi^{(2, \, n_2)} \, y \right] \, , \\
f_R^{(2, \, n_2, \, d_2)} (y, i) &=& \eta^{(d_2)}_i \sqrt{\dfrac{2}{\ell}} \, \sin \left[ m_\psi^{(2, \, n_2)} \, y \right] \, ,
\end{array}
\right.
\label{wave_funct_psi_1_rose}
\end{equation}
with the $N$-vector basis \eqref{basis_eta} so each mode propagates in only one petal. The $f_L^{(2, \, n_2, \, d_2)}$'s are discontinuous across the $V$-brane (except for $\mathcal{K}_N = \mathcal{R}_1$, the circle, where they can be taken continuous).
\end{itemize}

\subsubsection*{\boldmath \textcolor{black}{\textit{Second case}: $f_R^{(b, \, n_b, \, d_b)}(0,i) \neq 0$}}
The KK-mass spectrum is the same as in Eq.~\eqref{mass_spect_Psi_rose_bis}. Each KK-level $(b, n_b)$ is thus degenerate with the $N$ KK-modes of the level $(2, n_2)$ so $d_2 \in \llbracket 1, N+1 \rrbracket$: we label the KK-modes with nonvanishing wave functions at the junction by the triplet $(2, n_2, N+1)$. The KK-wave functions are
\begin{equation}
\left\{
\begin{array}{rcl}
f_L^{(2, \, n_2, \, N+1)} (y, i) &=& - \sqrt{\dfrac{2}{N \ell}} \, \sin \left[ m_\psi^{(2, n_2)} \, y \right] \, , \\
f_R^{(2, \, n_2, \, N+1)} (y, i) &=& \sqrt{\dfrac{2}{N \ell}} \, \cos \left[ m_\psi^{(2, n_2)} \, y \right] \, ,
\end{array}
\right.
\label{wave_func_psi_rose_period}
\end{equation}
where the $f_L^{(2, \, n_2, \, N+1)}$'s can be taken continuous across the $V$-brane and propagate in the whole rose graph.\\

Like for the scalar field, all KK-towers labeled by $b$ are present in the spectrum. Each excited KK-level is vector-like for both compactifications, and the mass gap between the KK-modes is of the order of $1/\ell$.

\section{A Low 5D Planck Scale with a Star/Rose Extra Dimension}
\label{ADD_star_rose}
In this section, we propose an ADD model with brane-localized 4D SM-fields where gravity propagates in a large star/rose EDS with large $N$ and a natural value for $\ell$.

\subsection{Lowering the Gravity Scale}
In order to include gravity in our setup, one has to extend general relativity with a star/rose EDS and thus to generalize the definitions of Subsection~\ref{Spacetime_Geometry} to the case of a curved geometry.
Consider $N$ identical 5D Lorentzian manifolds labeled by $i \in \llbracket 1, N \rrbracket$. Each of them is equipped with a coordinate system $z_i^M = (x^\mu, y, i)$, a metric $g_{i,MN}^{(5)}$, and has two timelike boundaries at $y=0, \ell$. One gets the star EDS by gluing the $N$ 5D spacetimes at one of their boundaries: the $J$-brane, and the $N$ remaining boundaries are still called $B_i$-branes. The rose EDS is obtained by gluing the $2N$ boundaries of these 5D spacetimes at a single timelike hypersurface: the $V$-brane. For both geometries, the 5D Lorentz-Poincaré transformations are local symmetries in the bulk, but they are explicitly broken to the 4D ones at the $J/B_i/V$-brane positions. These branes are rigid so that there are no branons associated to their fluctuations in the bulk. The line element in the $i^\text{th}$ leaf/petal is
\begin{equation}
ds_i^2 = g_{i,MN}^{(5)}dz^M_i dz^N_i \, .
\end{equation}
By construction, the metric is continuous across the $J/V$-brane, but this is not necessary the case for its derivatives. We suppose an exact permutation symmetry $\Sigma_N$ between the leaves/petals.\\

We are looking for a particle physics model where the background geometry is flat, so we consider a vanishing 5D cosmological constant and tensionless branes\footnote{When 4D fields are localized on a 3-brane, one can get an effective tensionless brane if the tension generated by the vacuum energy of the fields is exactly balanced by the bare tension of the brane.}. It is not the purpose of this article to solve the fine-tuning issues related to tensionless branes and a vanishing 5D cosmological constant. For the moment, we ignore the matter fields, the action is
\begin{equation}
S_{bulk} = - \dfrac{\left[ \Lambda_P^{(5)} \right]^3}{2} \int d^4x \sum_{i=1}^N \int_0^\ell dy \sqrt{\left|g_i^{(5)}\right|} \ R_i^{(5)} + \text{GH boundary terms} \, .
\label{EH5D}
\end{equation}
where $g_i^{(5)}$ is the determinant of the 5D metric $g_{i,MN}^{(5)}$ and $R_i^{(5)}$ is the 5D Ricci scalar in the $i^\text{th}$ leaf/petal. Gibbons-Hawking (GH) boundary terms \cite{Gibbons:1976ue} are added at both ends of the leaves/petals, i.e. on each side of the 3-branes, in order to have a well-defined action principle \cite{Chamblin:1999ya, Lalak:2001fd, Carena:2005gq}. We want to see how the 4D Einstein-Hilbert action:
\begin{equation}
S_{EH} = - \dfrac{\left[ \Lambda_P^{(4)} \right]^3}{2} \int d^4x \sqrt{\left|g^{(4)}\right|} \ R^{(4)} \, ,
\label{EH4D}
\end{equation}
is embedded into the 5D action \eqref{EH5D}, where $g^{(4)}$ is the determinant of the 4D metric $g_{\mu \nu}^{(4)}$ and $R^{(4)}$ is the 4D Ricci scalar. For that purpose, we consider the 4D metric fluctuations $h_{\mu \nu}(x^\mu)$ around a flat background:
\begin{equation}
ds_i^2 = \left( \eta_{\mu \nu} + h_{\mu \nu} \right) dx^\mu dx^\nu - dy^2 \, ,
\end{equation}
and we get
\begin{equation}
\left|g_i^{(5)}\right| = \left|g^{(4)}\right| \, , \ \ \ R_i^{(5)} = R^{(4)} \, .
\end{equation}
By integrating over the EDS in the 5D action \eqref{EH5D}, we recover the 4D action \eqref{EH4D} with the relation:
\begin{equation}
\left[\Lambda_P^{(4)}\right]^2 = L \, \left[ \Lambda_P^{(5)} \right]^{3} \, , \ \ \  L = N \ell \, .
\label{ADD_formula_star}
\end{equation}

To solve the gauge hierarchy problem with the hypothesis of an exact global $\Sigma_N$ symmetry (cf. Section~\ref{break_sym} for a discussion when this hypothesis is relaxed), we choose $\Lambda_P^{(5)} \simeq 1$ TeV, obtained with $L \simeq 1 \times 10^{12}$ m. In order to be in the EFT regime, i.e. below the 5D Planck scale, we need $\ell > \ell_P^{(5)}$, with $\ell_P^{(5)}=1/\Lambda_P^{(5)} \simeq 2 \times 10^{-19}$ m. In practice, $\ell/\ell_P^{(5)} \simeq 10$ with a large $N \simeq 6 \times 10^{29}$ should be enough, and thus a KK-mass scale near the EW-scale: $M_{KK} = 1/\ell \simeq 100$ GeV. Such heavy KK-gravitons evade completely the constraints from submillimeter tests of 4D gravitational Newtons's law, stellar physics and cosmology. If one allows for $1\%$ of fine-tuning for $m_h$ by pushing $\Lambda_P^{(5)}$ up to 10 TeV with $N \simeq 6 \times 10^{27}$, one can even allow for $M_{KK} \simeq 1$ TeV. Moreover, if the concepts of space and volume still make sense at the Planck scale $\Lambda_P^{(5)}$, by taking $\ell \simeq \ell_P^{(5)}$ and $N \simeq 6 \times 10^{30}$ to get $\Lambda_P^{(5)} \simeq 1$ TeV, there is no tower of KK-gravitons in the EFT, instead the first experimental hints for a low 5D Planck scale are strongly coupled quantum gravity phenomena near $\Lambda_P^{(5)}$.\\

Such a large $N$ seems puzzling at first glance: the reader would wonder whether our proposition is just a reformulation of the gauge hierarchy into why $N$ is large. However, in the EFT, $N$ is a conserved integer, so it is stable under radiative corrections, it does not need to be dynamically stabilized, and has no preferred value. When $\ell \gg \ell_P^{(5)}$, the models proposed in this article are formulated in the context of EFT's defined on a classical background spacetime $\mathcal{M}_4 \times \mathcal{K}_N$, where the number of leaves/petals $N$ is fixed by the definition of the model, even in presence of gravity. Possibly, $N$ becomes a dynamical quantity in a theory of Planckian gravity involving a quantum spacetime. The situation is somewhat similar to other beyond the SM-scenarii in the literature to solve the hierarchy problem where a large conserved integer is involved (cf. Refs.~\cite{Kaloper:2000jb, ArkaniHamed:1998kx, Corley:2001rt, Dvali:2007hz,Dvali:2007wp,Dvali:2008fd,Dvali:2008jb,Dvali:2009fw,Dvali:2009ne,Dvali:2019ewm, Arkani-Hamed:2016rle}).\\

The only quantity which needs to be dynamically stabilized is the leaf (petal) length (circumference) $\ell$, otherwise the radion, i.e. the scalar field which represents the fluctuations of $\ell$, remains massless and conflicts with the null result from the search for a new force of infinite range. Moreover, if only the graviton propagates into the EDS, the bosonic quantum loops are known to make the EDS unstable, which shrinks to a point \cite{Appelquist:1982zs, Appelquist:1983vs}. On the one hand, if $\ell \gtrsim \mathcal{O}(10) \times \ell_P^{(5)}$, the corrections of Planckian gravity can safely be neglected (EFT regime), and subsequently we assume an additional mechanism which stabilizes $\ell$. On the other hand, if $\ell \sim \ell_P^{(5)}$, one expects $\mathcal{O}(1)$ corrections from Planckian gravity, and one needs a complete theory of gravity to formulate the model and to address its stabilization.\\
 
In the EFT regime, we have supposed the existence of an exact global $\Sigma_N$ symmetry of the $N$-star/rose, which is not realistic since gravity is supposed to break global symmetries (cf. Section~\ref{break_sym} for a discussion on the impact on the resolution of the gauge hierarchy problem). This is reminiscent of the 5D models with a universal extra dimension (UED) compactified on an interval symmetric under an exact $\mathbb{Z}_2$ reflection with respect to its midpoint \cite{Appelquist:2000nn}. One can also mention the model of Ref.~\cite{Agashe:2007jb} where two identical slices of AdS$_5$ are glued to a common UV-brane. The EDS of these models can be stabilized by the dynamics of additional bulk fields at the quantum level by a balance between the contribution of bosonic and fermionic loops \cite{Ponton:2001hq}, or at the classical level by a potential for a scalar field as in the Goldberger-Wise (GW) mechanism originally proposed for a warped EDS \cite{Goldberger:1999uk, DeWolfe:1999cp, Csaki:2000zn}. In Ref.~\cite{Law:2010pv}, it is shown how to stabilize a $N$-star $\mathcal{S}_N$ with the potential of a different scalar field in each leaf, these $N$ scalar fields are related by the $\Sigma_N$ symmetry. One can apply these mechanisms here with $N$ large. Other stabilization mechanisms were proposed in Ref.~\cite{ArkaniHamed:1998kx}, in particular it is possible to stabilize an EDS compactified on a circle with the help of a complex scalar field with a topologically conserved winding number. As it is possible to stabilize one petal by this mechanism, one can repeat it with a different scalar field in each petal of the $N$-rose $\mathcal{R}_N$. Both for $\mathcal{S}_N$ and $\mathcal{R}_N$, the $N$ scalar fields meet only at the junction $J/V$, and we assume that they interact only through gravity, like the $N$ copies of the SM in Ref.~\cite{Dvali:2007wp, Dvali:2009ne}. Therefore, the picture reduces to stabilize $N$ independent leaves/petals, which is a simplified version of the mechanism in Ref.~\cite{Law:2010pv}. If the bulk fields in all leaves/petals have an exact global $\Sigma_N$ symmetry, the geometrical $\Sigma_N$ symmetry is preserved (cf. Section~\ref{break_sym} for a discussion when this is not the case).

\subsection{Embedding the Standard Model Fields}
Up to this subsection, we did not mention how we embed the SM-fields into the proposed spacetime geometries. If we do not want to break explicitly the $\Sigma_N$ symmetry with the SM-fields, one has two possibilities:
\begin{itemize} 
\item The SM-fields are 5D fields propagating into the whole star/rose graph.
\item The SM-fields are 4D fields localized on the $J/V$-brane.
\end{itemize}

For 5D SM-fields, the star graph is the only one which allows one chiral zero-mode for each 5D fermion, which is essential to embed the chiral structure of the EW sector of the SM with a large $N$. Moreover, one should choose the KK-mass scale at least at $M_{KK} = 1/\ell \sim \mathcal{O}(1)$ TeV, since we have not discovered KK-excitations of the SM-particles yet. However, one should also consider the magnitude of the couplings of the zero-mode gauge bosons. In a model with a 5D gauge field propagating on a flat star/rose EDS, we have to determine how the higher dimensional gauge coupling $g_5$ (of mass dimension $-1/2$) is related to the zero-mode gauge coupling $g_4$. The action for a 5D abelian gauge field $A_M$ propagating into the whole star/rose EDS is\footnote{Only the kinetic term of the gauge field is important for our argument, so it can be applied also to nonabelian gauge fields.}
\begin{equation}
S^{(5)}_{gauge} = - \dfrac{1}{4 g_5^2} \int d^4x \int_0^\ell dy \ \eta_{i,}^{\ MP} \eta_{i,}^{\ NS} F_{MN} F_{PS}
\label{5D_gauge}
\end{equation}
with $F_{MN} = \partial_M A_N - \partial_N A_M$. Consider only the zero-mode whose wave function is flat: $A_M (x^\mu, y, i) = A_\mu^{(0)}(x^\mu)$. If we integrate over the EDS in the 5D action \eqref{5D_gauge}, we find that the effective 4D action include the action of a 4D abelian gauge field $A_\mu^{(0)}$:
\begin{equation}
S^{(4)}_{gauge} = - \dfrac{1}{4 g_4^2} \int d^4x \int_0^\ell dy \ \eta^{\mu \rho} \eta^{\nu \sigma} F^{(0)}_{\mu \nu} F^{(0)}_{\rho \sigma}
\end{equation}
with $F^{(0)}_{\mu \nu} = \partial_\mu A_\nu^{(0)} - \partial_\nu A_\mu^{(0)}$ and the relation:
\begin{equation}
g_4 = \dfrac{g_5}{\sqrt{L}} \, .
\label{gauge_ADD}
\end{equation}
With a natural value for the gauge coupling
\begin{equation}
g_5 \sim \sqrt{\ell_P^{(5)}} \, ,
\end{equation}
one obtains with Eq.~\eqref{ADD_formula}:
\begin{equation}
g_4 \sim \sqrt{\dfrac{\ell_P^{(5)}}{L}} = \dfrac{\Lambda_P^{(5)}}{\Lambda_P^{(4)}} \sim 10^{-16} \, ,
\end{equation}
so $g_4$ is a very tiny coupling and cannot be identified with a SM-gauge coupling. This result depends only on the volume of the compactified space, i.e. on the hierarchy between the 4D and 5D Planck scales. After this discussion, it is clear that in the case of a star/rose EDS with a large $N$, the SM-fields must be localized on a 3-brane, like in the other ADD models in the literature.\\

Consider a 5D EFT with a brane. The cut-off in the bulk and the 3-brane thickness are noted $\Lambda$ and $\epsilon$ respectively. There are two cases~\cite{delAguila:2006atw}:

\begin{itemize}
\item The fat brane ($\epsilon > 1/\Lambda$): its microscopic description is in the range of validity of the 5D EFT. Usually, a fat brane is a topological defect \cite{Akama:1982jy, Rubakov:1983bb} and it is necessary to provide a field theoretical mechanism to trap the zero-modes of 5D fields of various spins in the neighborhood of the brane \cite{Visser:1985qm, Jackiw:1975fn, Dvali:1996xe, Dubovsky:2001pe, Ohta:2010fu}. The topological defects are the first prototypes of braneworlds in the literature and are chosen by ADD to trap the SM-fields in their first article on LED's \cite{ArkaniHamed:1998rs}. It is also possible to localize the zero-modes of 5D fields towards orbifold fixed points or spacetime boundaries with large 5D masses of brane-localized kinetic terms \cite{Dvali:2000rx, Dubovsky:2001pe, Fichet:2019owx}. Irrespectively of the trapping mechanism of the fat brane, we speak about quasi-localized 5D fields.
\item The thin brane ($\epsilon \leq 1/\Lambda$): its microscopic description is outside the range of validity of the 5D EFT. A thin brane is described in the EFT by an infinitely thin hypersurface where 4D fields are strictly localized \cite{Sundrum:1998sj, Csaki:2004ay, Fichet:2019owx}. The trapping mechanism of the fields is relegated to the UV-completion. This case became popular when it was realized that 4D fields can live in the worldvolume of solitonic objects in some UV-completions, like D-brane stacks in superstring theories where matter fields are described by open strings attached to them \cite{Polchinski:1996na, Bachas:1998rg, Johnson:2003gi}. In EFT's, orbifold fixed points, spacetime boundaries or metric graph vertices are perfect candidates for thin branes. One can also obtain a thin brane by integrating out the width of a fat brane: one gets an EFT with a cut-off equal to the inverse of the brane width, and 4D fields (the zero-modes of the quasi-localized 5D fields of the UV-completion) strictly localized on the thin brane (depending on the quasi-localization mechanism, the excited KK-modes do not necessarily decouple \cite{Fichet:2019owx}). Quickly, after the theoretical discovery of D-branes, physicists explored the new possibilities offered by thin branes \cite{Antoniadis:1998ig, ArkaniHamed:1998nn, Randall:1999ee, Randall:1999vf, Lykken:1999nb, Kogan:1999wc, Gregory:2000jc, Dvali:2000hr}. 
\end{itemize}

For the models studied in this article, if one considers quasi-localized 5D SM-fields on the $J/V$-branes (fat branes), one has a problem. Indeed, consider the $N$-star $\mathcal{S}_N$, the fat brane has a thickness $\epsilon > \ell_P^{(5)}$ extended into each leaf, so a zero-mode gauge coupling $g_4$ is related to the 5D gauge coupling $g_5 \sim \sqrt{\ell_P^{(5)}}$ as
\begin{equation}
g_4 \sim \dfrac{g_5}{\sqrt{N \epsilon}} \lesssim \mathcal{O} \left( \dfrac{1}{\sqrt{N}} \right) \, ,
\label{gauge_fat_J}
\end{equation}
similar to Eq.~\eqref{gauge_ADD}. As $N$ is large, the model will suffer from the same problem as for bulk SM-fields: the gauge couplings of the zero-modes are too suppressed to match the values measured in experiments. The same issue arises with a fat $V$-brane in the case of the $N$-rose $\mathcal{R}_N$.
Therefore, we will consider only 4D SM-fields localized on a thin $J/V$-brane: the SM-gauge fields do not arise from the limit of quasi-localized 5D fields which propagate into the leaves/petals, so the gauge couplings are not suppressed by $\sqrt{N}$. However, as already discussed, in a UV-completion including gravity, the singular behavior of the junction should be regularized, for example by integrating in $q-1$ new transverse dimensions as in the case of a $q$D graph-like manifold. Concerning the UV-origin of the brane-localized 4D SM-fields, one can imagine a UV-completion with a $q$D graph-like manifold and a fat $J/V$-brane made of $q$D quasi-localized fields. However, the wave functions of the zero-modes must be highly peaked inside the protrusion \cite{Kuchment2002} at the vertex $J/V$, i.e. they must decrease quickly inside the vertex protrusion such that they are suppressed at least by $1/\sqrt{N}$ at the entrance of a leaf/petal to avoid the problem of Eq.~\eqref{gauge_fat_J}. One can also imagine another UV-completion: if it is possible to generate graph-like 6D manifolds in superstring theories, the 4D SM-fields may live in the worldvolume of a D-brane stack into the protrusion of the $J/V$ vertex.

\subsection{Phenomenology}
\label{pheno_star_rose_graviton}

\subsubsection{Kaluza-Klein Gravitons}
\label{KK-graviton}
In an ADD model, gravity and possibly other exotic fields propagate into the EDS's. It is crucial for our proposition to have an idea of the implication of gravitons propagating into the bulk. The KK-dimensional reduction of a 5D graviton \cite{Csaki:2004ay} leads to a tower of KK-gravitons with a zero-mode, and one massless graviscalar (the radion). A massless graviphoton is also present if there is no boundary for the EDS (present for the $N$-rose $\mathcal{R}_N$ and absent for the $N$-star $\mathcal{S}_N$). As the existence of KK-gravitons can have important phenomenological effects, one has to extract their KK-mass spectrum and their couplings to the SM-fields. By a suitable gauge choice, the Euler-Lagrange equations for a 5D massless graviton reduce to Klein-Gordon equations. As it was suggested in Ref.~\cite{Kim:2005aa}, one can thus study a 5D massless real scalar field coupled minimally to the energy momentum tensor of the SM to obtain the KK-mass spectrum and the couplings of spinless KK-gravitons. This 5D spinless graviton $\Phi$ couples to the energy momentum sources through the effective induced metric on the $J$-brane:
\begin{equation}
g_{\mu \nu}^{(5)} (x^\mu, 0, i) = \left( 1 + \dfrac{1}{2\left[ \Lambda_P^{(5)} \right]^{3/2}} \, \Phi (x^\mu, 0, i) \right) \eta_{\mu \nu} \, .
\end{equation}
Remember that the 5D metric has to be continuous at the junction, so it is also the case of $\Phi$ in our toy model of spinless gravitons. Moreover, $\Phi$ is a 5D massless real scalar field, so the KK-mode analysis is the same as in Subsection~\ref{KK_scalar_20} with $M_\Phi = 0$. We treat the brane-localized interactions with the SM-fields as a perturbation. The zero-mode is identified with the 4D massless graviton, it propagates into the whole star/rose graph and its wave function is given in Eq.~\ref{zero_mode_sol_Phi_2}. We focus on the compactification on $\mathcal{S}_N$ since the case of $\mathcal{R}_N$ is very similar. The coupling of the 5D spinless graviton to the energy momentum tensor 
\begin{equation}
T^{\mu \nu} = \left. \dfrac{2}{\sqrt{|g|}} \dfrac{\delta S_{SM}}{\delta g_{\mu \nu}} \right|_{g_{\mu\nu} = \eta_{\mu\nu}}
\end{equation}
of the 4D SM-fields (of action $S_{SM}$) localized on the $J$-brane is
\begin{equation}
\left(\dfrac{1}{2\left[ \Lambda_P^{(5)} \right]^{3/2}} \, \widetilde{\Phi} \, T^\mu_\mu \, \delta_J\right)[1] = \int d^4x \sum_{i=1}^N \ \dfrac{1}{2N\left[ \Lambda_P^{(5)} \right]^{3/2}} \,  \Phi (x^\mu, 0, i) \, T^\mu_\mu \, .
\label{int_Phi-SM_star_1}
\end{equation}
We note $n_*$ the number of KK-modes which couple to the SM-fields below the cut-off $\Lambda_P^{(5)}$. Only the KK-gravitons $b=2$ with a wave function which does not vanish on the $J$-brane couple to the SM-fields (see the wave functions of the KK-modes in Subsection~\ref{KK_scalar_20}). The interaction term \eqref{int_Phi-SM_star_1} gives
\begin{equation}
\int d^4x \left[ \dfrac{1}{2\Lambda_P^{(4)}} \, \phi^{(0, \, 0, \, 1)} (x^\mu) \, T^\mu_\mu + \dfrac{\sqrt{2}}{2\Lambda_P^{(4)}} \, \sum_{n_2 = 1}^{n_*} (-1)^{n_2} \, \phi^{(2, \, n_2, \, 1)}(x^\mu) \, T^\mu_\mu \right] \, ,
\label{int_Phi-SM_star_2}
\end{equation}
where
\begin{equation}
n_* \sim \dfrac{\Lambda_P^{(5)} \, \ell}{\pi} \, .
\label{n_max}
\end{equation}
The KK-modes whose wave functions do not vanish at $y=0$ couple individually to the energy momentum tensor of the SM with a coupling suppressed by $\Lambda_P^{(4)}$: they are thus very feebly coupled, and the probability $P_1$ to emit a single KK-graviton is proportional to its coupling squared:
\begin{equation}
P_1 \propto \left[ \dfrac{E}{\Lambda_P^{(4)}} \right]^2 \, ,
\label{P_1}
\end{equation}
where $E$ is the energy of matter originating from $T_\mu^\mu$ in Eq.~\eqref{int_Phi-SM_star_2}. We compare two benchmark scenarii with $\Lambda_P^{(5)} \simeq 1$ TeV:

\paragraph{Benchmark scenario \#1 --}
We take $N = 1$. This case is the traditional situation of ADD models in the literature with only one leaf. From Eq.~\eqref{ADD_formula_star}, we have $M_{KK} = 1/\ell \sim \mathcal{O}(10^{-18})$ eV, which is excluded by the success of 4D gravitational Newton's law at the scale of the solar system. Eqs.~\eqref{ADD_formula_star} and \eqref{n_max} give
\begin{equation}
n_* \sim \dfrac{1}{\pi} \left[ \dfrac{\Lambda_P^{(4)}}{\Lambda_P^{(5)}} \right]^2 \sim 10^{30} \, ,
\end{equation}
so we have a large number of KK-gravitons below the cut-off. At colliders with a center of mass energy which reaches $\Lambda_P^{(5)}$, the probability to produce one out of $n_*$ gravitons becomes then
\begin{equation}
P_* = n_* \, P_1 \propto \left[ \dfrac{E}{\Lambda_P^{(5)}} \right]^2 \, .
\label{P*ADD}
\end{equation}
where we used Eqs.~\eqref{ADD_formula_star} , \eqref{n_max} and \eqref{P_1}.
This last result is also valid in more realistic models with more than one EDS, which can pass with success the submillimeter tests of 4D gravitational Newton's law. The KK-tower can thus be probed and constrained at the Large Hadron Collider (LHC, $\sqrt{s}=13$~TeV) provided $E$ reaches $\Lambda_P^{(5)}$.

\paragraph{Benchmark scenario \#2 --} We take $N \simeq 6 \times 10^{29}$. In this case, the large volume in Eq.~\eqref{ADD_formula_star} is generated by a large $N$ and $M_{KK} = 1/\ell \simeq 100$ GeV. Thus, there are few KK-modes which couple to the SM-fields: $n_* \simeq 3$ from Eq.~\eqref{n_max}, and $P_* \sim P_1$ at the LHC. So the KK-tower is completely invisible in current experiments. The compactification on $\mathcal{S}_N$ can thus circumvent the current LHC constraints on the KK-gravitons of traditional ADD models.\\

However, these results follow from the zero-thickness brane hypothesis. How are they modified by a brane width in the UV-completion? Indeed, we have already discussed that one expects that the singular behavior of the junction is softened in a UV-completion including gravity. After integrating out the UV-degrees of freedom, one is left with an effective brane form factor as in Ref.~\cite{Kiritsis:2001bc} to model the brane width\footnote{We stress that this effective brane form factor has nothing to do with the wave function of the zero-mode of quasi-localized 5D fields on the $J$-brane but is related to the UV-description of the brane. We have already discussed that the brane-localized SM-fields are 4D degrees of freedom in the EFT.}. It is a function $\mathcal{B}_J(y)$ rapidly decreasing over a distance $\ell_P^{(5)}$ and normalized such that
\begin{equation}
\sum_{i=1}^N \int_0^\ell dy \ \mathcal{B}_J(y,i) = 1 \, .
\end{equation}
One can perform a moment expansion of $\mathcal{B}_J(y)=\Lambda_P^{(5)} b \left( \Lambda_P^{(5)} y \right)$, where $b(y)$ is an intermediate function defined for convenience, such that
\begin{equation}
\widetilde{\mathcal{B}_J} =  \sum_{n=0}^{+ \infty} \dfrac{b_n}{\left[\Lambda_P^{(5)}\right]^{n}} \, \partial_y^n \delta_J \, ,
\label{asymp_exp}
\end{equation}
with
\begin{equation}
b_n = \dfrac{(-1)^n}{n!} \int_0^\ell dy \ y^n \, b(y) \, .
\end{equation}
The action describing the interaction between the spinless graviton and the SM-fields is
\begin{align}
&\int d^4x \left( \dfrac{1}{2 \left[ \Lambda_P^{(5)} \right]^{n + 3/2}} \, \widetilde{\Phi} \, T^\mu_\mu \, \widetilde{\mathcal{B}_J} \right) [\mathbf{1}] \nonumber \\
&=\int d^4x \ \left(\sum_{n=1}^{+\infty}\dfrac{b_n}{2 \left[ \Lambda_P^{(5)} \right]^{n + 3/2}} \, \widetilde{\Phi} \, T^\mu_\mu \, \partial_y^n \delta_J\right)[\mathbf{1}] \nonumber \\
&= \int d^4x \sum_{n=1}^{+\infty} \sum_{i=1}^N \ \dfrac{(-1)^n \, b_n}{2N\left[ \Lambda_P^{(5)} \right]^{n+3/2}} \, \partial_y^n \Phi (x^\mu, 0, i) \, T^\mu_\mu \, .
\end{align}
One can naively think that the large number of KK-gravitons which do not couple to the SM-fields through the operator \eqref{int_Phi-SM_star_1} will have nonvanishing couplings to the SM via the higher-dimensional operators. Then, one expects that $P_*$ is less suppressed than in Eq.~\eqref{P_1}. However, this is not the case. Indeed, by using the equations for the wave functions \eqref{wave_eq_Phi_star}, one can show that
\begin{equation}
\forall l \geq 1 \left\{
\begin{array}{l c l}
\partial_y^{2l} f_\phi^{(b, \, n_b, \, d_b)} (0, i) &=& (-1)^l \, \left[ k_\phi^{(b, \, n_b)} \right]^{2l} f_\phi^{(b, \, n_b, \, d_b)} (0, i) \, , \\ \\
\partial_y^{2l+1} f_\phi^{(b, \, n_b, \, d_b)} (0, i) &=& (-1)^l \, \left[ k_\phi^{(b, \, n_b)} \right]^{2l} \partial_y f_\phi^{(b, \, n_b, \, d_b)} (0, i) \, .
\end{array}
\right.
\end{equation}
Therefore, for $n$ even, again only the tower $b=2$ contributes, with an extra suppression factor $\left[ M_{KK}/\Lambda_P^{(5)} \right]^n$. For $n$ odd, the Neumann-Kirchhoff junction conditions \eqref{JC_Phi_star} imply that these operators vanish. We conclude that even when we take into account the brane width in the UV, the KK-graviton towers are still invisible at the LHC ($\sqrt{s}=14$~TeV). The KK-gravitons with $b \neq 2$ constitute a hidden sector.

\subsubsection{Ultraviolet Gravitational Objects}
Black holes are expected to appear near the cutoff scale $\Lambda_P^{(5)}$, when the coupling to 5D gravitons becomes nonperturbative. However, in the case of the benchmark scenario \#2, we saw that the coupling of the energy-momentum tensor of the SM to the linear superposition of KK-gravitons is suppressed by $\Lambda_P^{(4)}$ instead of $\Lambda_P^{(5)}$, so one expects that the couplings of the brane-localized SM-fields to the tower of KK-gravitons remains perturbative well above $\Lambda_P^{(5)}$, questioning the possibility of producing black holes in trans-Planckian collisions of SM-particles. However, once the linear superposition of KK-gravitons with a trans-Planckian energy leaves the $J/V$-brane, where it was perturbatively produced through SM-fields in a trans-Planckian collision, it will interact with all the KK-gravitons, including those whose wave functions vanish on the $J/V$-brane. This last process is nonperturbative above $\Lambda_P^{(5)}$ and will produce a black hole. Near this threshold, the black holes are dominated by quantum corrections, we speak about quantum black holes (QBH's) \cite{Rizzo:2006zb, Alberghi:2006km, Meade:2007sz, Casadio:2008qy, Calmet:2008dg, Gingrich:2009hj, Gingrich:2010ed, Dvali:2010gv, Nicolini:2011nz, Mureika:2011hg, Calmet:2012fv, Kilic:2012wp, Belyaev:2014ljc, Arsene:2016kvf} which need a complete theory of quantum gravity to be described. Besides, the lightest QBH, the Planckion \cite{Treder:1985kb, Dvali:2016ovn}, is the last stage of the evaporation of a semi-classical black hole by Hawking radiation. In some models, this black hole remnant \cite{Koch:2005ks, Dvali:2010gv, Bellagamba:2012wz, Alberghi:2013hca} is stable and one can speculate that it can constitute a part of dark matter \cite{Conley:2006jg, Dvali:2010gv, Nakama:2018lwy}. There are also a large number of KK-gravitons below the TeV-scale whose wave functions vanish on the $J/V$-brane, where the SM-fields are localized (cf. Subsection~\ref{KK_scalar_20}): these KK-gravitons interact only with gravity in the bulk, and constitute a natural hidden sector which could be populated by black hole evaporation during the early Universe.

\subsection{$\Sigma_N$ symmetry breaking by gravity}
\label{break_sym}
Our effective model (as well as the traditional UED models) has a global $\Sigma_N$ symmetry which acts on the geometry. One of the Swampland conjectures is the absence of global symmetries in a complete theory of quantum gravity (cf. Refs.~\cite{Brennan:2017rbf, Palti:2019pca} for reviews on the Swampland program). If true, there is no global $\Sigma_N$ symmetry and one has a different leaf/petal size $\ell_i$ for each $i$. One can define the average of the leaf/petal size as
\begin{equation}
\langle \ell \rangle = \dfrac{1}{N} \sum_i \ell_i \, ,
\end{equation}
such as Eq.~\eqref{ADD_formula_star} is still valid but with $L = N \langle \ell \rangle$. The mass spectrum must be studied numerically. In general, the mass scale of the lightest KK-mode is given by the inverse of the largest $\ell_i$. If the $\ell_i$'s are incommensurate (i.e. $\forall i \neq j, \ \ell_i/\ell_j \notin \mathbb{Q}$), the KK-spectrum is chaotic \cite{PhysRevLett.79.4794, Kottos_1999, Gaspard2000a, Dabaghian2001a, PhysRevLett.88.044101, kottos2005quantum, Cacciapaglia:2006tg}. Moreover, there are more KK-modes below the cut-off $\Lambda_P^{(5)}$ which couple to the $J$-brane for different $\ell_i$'s: $P_*$ is thus less suppressed than for identical $\ell_i$'s such that one is in an intermediate situation between the benchmark scenarii \#1 and \#2. If the $\ell_i$'s are incommensurate, there are no KK-modes whose wave functions vanish at the $J$-brane and the KK-spectrum is chaotic.\\

In a realistic model including gravitational effects, instead of an exact global $\Sigma_N$ symmetry, one should consider a geometry with an approximate one. In case of GW mechanisms within each leaf, the $\ell_i$'s would depend on the mass parameter for the GW scalar field within each leaf. Identical leaf lengths (exact $\Sigma_N$ symmetry) corresponds to $\Sigma_N$ symmetric mass parameters within all leafs. The question is then in how far quantum gravity effects affect classically $\Sigma_N$ symmetric mass parameters. If these modifications remain within $\sim 10\%$ for all leafs, i.e. the effects of gravity can be considered as a perturbation of the geometry with an exact $\Sigma_N$ symmetry, the hierarchy problem could be considered as solved, with the common scale being given by $1/\langle \ell \rangle$. On the other hand, if the modified leaf lengths follow a statistical distribution (like a Gaussian) around a central value $\langle \ell \rangle$, the hierarchy problem can be considered as solved only if this distribution is extremely narrow such that the number of very light KK-gravitons remains compatible with present constraints, cf. Subsection~\ref{KK-graviton}. In the absence of a concrete theory of quantum gravity, it is impossible to answer these questions. Therefore, the toy model of this article with an exact global $\Sigma_N$ symmetry should not be considered as a viable solution of the gauge hierarchy problem but as a scenario within which (among others) a solution of the hierarchy problem may be possible.

\section{Toy Model of Small Dirac Neutrino Masses}
\label{Dirac_Neutrinos}

\subsection{Zero-Mode Approximation}
It is known from Refs.~\cite{Dienes:1998sb, ArkaniHamed:1998vp, Dvali:1999cn} that if the left-handed neutrinos, localized on the SM-brane, interact with gauge singlet neutrinos, propagating in the bulk in the form of an internal torus $\left( \mathcal{R}_1 \right)^q$ of radius $R$ and large volume $\mathcal{V}_q$, one can get small Dirac masses for the neutrinos. We want to see if it is possible to build such a model for a LED compactified on the metric graph $\mathcal{K}_N$. For the compactification on a star/rose graph, one takes a 4D left-handed neutrino $\nu_L$ of mass dimension 3/2 localized on the $J/V$-brane, and a 5D gauge singlet neutrino $\Psi$ of mass dimension 2 propagating into the bulk. The action of the model is
\begin{align}
S_{\nu} &= S_\Psi + \int d^4x \left( \widetilde{\mathcal{L}_\nu} + \widetilde{\mathcal{L}_{\Psi \nu}} \right) \delta_{J/V}[\mathbf{1}] \, , \nonumber \\
&= S_\Psi + \int d^4x \left( \mathcal{L}_\nu + \left. \widetilde{\mathcal{L}_{\Psi \nu}} \right|_{y=0} \right) \, .
\label{S_nu}
\end{align}
The free action $S_\Psi$ is given by Eq.~\eqref{S_Psi_star}, and
\begin{equation}
\mathcal{L}_\nu = \dfrac{i}{2} \, \nu_L^\dagger \bar{\sigma}^\mu \overleftrightarrow{\partial_\mu} \nu_L \, .
\end{equation}
The brane-localized mass term is
\begin{equation}
\widetilde{\mathcal{L}_{\psi \nu}} = - \dfrac{y_\nu^{(5)} v}{\sqrt{2}} \, \nu_L^\dagger \widetilde{\Psi_R} + \rm{H.c.} \, ,
\label{nu_Yuk}
\end{equation}
where $v$ is the Higgs VEV and $y_\nu^{(5)}$ is the 5D Yukawa coupling of mass dimension $-1/2$. $y_\nu^{(5)}$ can be taken real, since a phase shift of the Yukawa coupling can be compensated by a phase shift of the field $\nu_L$. We have imposed that the leptonic number $L$ is conserved, so $U(1)_L$ is a symmetry of the model: accordingly, bulk and brane-localized Majorana mass terms for the neutrino fields are not allowed. We have also assumed the absence of a bulk Dirac mass term to simplify the discussion. By adopting a perturbative approach, where $\mathcal{L}_{\psi \nu}$ is treated as a perturbation, one can perform the KK-dimensional reduction of Section~\ref{Dirac_field} where we have imposed the continuity on $\Psi_R$ at the junction (as we want only one zero-mode for $\Psi_R$ which propagates into the whole EDS) and $\Psi_L$ is allowed to be discontinuous (we will see in the exact treatment of Subsection~\ref{exact_treatment} that the brane-localized mass term induces a jump at the junction for $\Psi_L$). In the regime where we can use the zero-mode approximation, i.e. when one can neglect the mixing between the zero-mode and the KK-excitations:
\begin{equation}
\dfrac{y_\nu^{(5)} v}{\sqrt{2L}} \ll M_{KK} \equiv \dfrac{1}{\ell} \, ,
\end{equation}
we get a mass term for the zero-mode neutrino:
\begin{equation}
-m_\nu \, \nu_L^\dagger \psi_R^{(0, \, 0, \, 1)} + \rm{H.c.} \, ,
\end{equation}
with
\begin{equation}
m_\nu = \dfrac{y_\nu^{(5)} v}{\sqrt{2}} \, f_R^{(0, \, 0, \, 1)}(0)
= \dfrac{y_\nu^{(5)} v}{\sqrt{2L}} \, .
\label{m_nu_2}
\end{equation}
For a natural value
\begin{equation}
y_\nu^{(5)} \sim \sqrt{\ell_P^{(5)}} \, ,
\end{equation}
with $\Lambda_P^{(5)} \simeq 1 \ \rm{TeV}$, one has, from Eqs.~\eqref{ADD_formula_star} and \eqref{m_nu_2},
\begin{equation}
m_\nu \sim \dfrac{v}{\sqrt{2L \Lambda_P^{(5)}}} = \dfrac{v \Lambda_P^{(5)}}{\sqrt{2}\Lambda_P^{(4)}} \sim \mathcal{O}(0.1) \ \rm{meV} \, ,
\end{equation}
which is a good order of magnitude for the neutrino masses. As $m_\nu \ll M_{KK} \equiv 1/\ell$ for the benchmark scenario \#2, the zero-mode approximation is thus justified.\\

However, within the perturbative approach, we find that the $N$ left-handed zero-modes in Section~\ref{Dirac_field} for the $N$-rose $\mathcal{R}_N$ do not get masses from the brane-localized mass term \eqref{nu_Yuk}. They remain massless and do not mix with the left-handed neutrino localized on the $V$-brane: they are sterile neutrinos which do not participate in neutrino oscillations. However, they are coupled to gravity and should have an impact on the cosmological history. As our model requires a large $N$, it appears to be ruled out by cosmological constraints, which are sensitive to the number of light fermionic degrees of freedom. Even with a brane-localized reheaton, which does not couple to the modes with discontinuous wave functions like the $N$ left-handed zero-modes, mini-black hole evaporation should produce them in the early Universe. A solution could be to add a new ingredient to the model to give a mass to these $N$ zero-modes. A priori, our toy model is thus interesting only for the compactification on a star graph.

\subsection{Exact Treatment}
\label{exact_treatment}

\subsubsection{Euler-Lagrange Equations \& Junction/Boundary Conditions}
In this subsection, we take the effect of the brane-localized mass term $\widetilde{\mathcal{L}_{\Psi \nu}}$ on the KK-mass spectrum exactly into account with the 5D method of Ref.~\cite{Angelescu:2019viv, Nortier:2020xms}. From Hamilton's principle applied to the action $S_\nu$ \eqref{S_nu}, we get the Dirac-Weyl equations: Eq.~\eqref{Dirac_Psi_star} and
\begin{equation}
i \bar{\sigma}^\mu \partial_\mu \nu_L(x^\mu) - M \, \Psi_R(x^\mu, 0, i) = 0 \, ,
\label{nu_ELE_5D}
\end{equation}
with
\begin{equation}
M = \dfrac{y^{(5)}_\nu v}{\sqrt{2}} \, .
\end{equation}
We get also a Kirchhoff junction condition for the left-handed field on the $J/V$-brane:
\begin{equation}
\left\{
\begin{array}{rcl}
\displaystyle{\sum_{i=1}^N \Psi_L (x^\mu, 0, i) = M \, \nu_L(x^\mu)} & \text{for} & \mathcal{K}_N = \mathcal{S}_N \, ,   \\
\displaystyle{\sum_{i=1}^N \left[ \Psi_L (x^\mu, y, i) \right]_{y=0}^{\ell} = M \, \nu_L(x^\mu)} & \text{for} & \mathcal{K}_N = \mathcal{R}_N \, ,
\end{array}
\right.
\label{Kir_junc_mix_Psi}
\end{equation}
and Dirichlet boundary conditions \eqref{D_psi_B} for the left-handed field on the $B_i$-branes.

\subsubsection{Separation of Variables}
We want to solve the field equations by separation of variables and sum over all linearly independent solutions. We thus write the KK-decomposition \eqref{KK_Psi_star_1} and expand $\nu_L$ as a linear superposition of the left-handed KK-modes, which are mass eigenstates:
\begin{equation}
\nu_L(x^\mu) = \sum_{b} \ \sum_{n_b} \sum_{d_b} a^{(b, \, n_b, \, d_b)} \, \psi_L^{(b, \, n_b, \, d_b)} \left( x^\mu \right) \, ,
\label{nu_L_sep}
\end{equation}
with $a^{(b, \, n_b, \, d_b)} \in \mathbb{C}$. Indeed, the brane-localized mass term induces a mixing between the field $\nu_L$ and the KK-modes of $\Psi_L$ obtained in Section~\ref{Dirac_field}. Here, we expand the fields $\nu_L$ and $\Psi_L$ in the same KK-basis spanned by the $\psi_L^{(b, \, n_b, \, d_b)}$'s (the basis of the mass eigenstates). The reader can follow the discussion between Eqs.~\eqref{KK_Psi_star_1} and \eqref{orthonorm_KK-Phi_star}, we will use the same notations, but we stress that, in Section~\ref{Dirac_field}, $\psi_L^{(b, \, n_b, \, d_b)}$ is an element of the KK-basis without brane-localized mass term but here it is an element of the KK-basis including the effect of the brane-localized mass term. Besides, the orthonormalization conditions \eqref{orthonorm_KK-Phi_star} for the functions $f_{L/R}^{(b, \, n_b, \, d_b)} \neq 0$ are replaced by
\begin{equation}
\left\{
\begin{array}{rcl}
\displaystyle{\left[a^{(b, \, n_b, \, d_b)}\right]^* a^{(b', \, n'_{b'}, \, d'_{b'})} + \sum_{i=1}^N \int_0^\ell dy \ \left[ f_{L}^{(b, \, n_b, \, d_b)}(y, i) \right]^* \, f_{L}^{(b', \, n'_{b'}, \, d'_{b'})}(y, i)} &=& \delta^{bb'} \, \delta^{n_{b} n'_{b'}} \, \delta^{d_{b} d'_{b'}} \, , \\
\displaystyle{\sum_{i=1}^N \int_0^\ell dy \ \left[ f_{R}^{(b, \, n_b, \, d_b)}(y, i) \right]^* \, f_{R}^{(b', \, n'_{b'}, \, d'_{b'})}(y, i)} &=& \delta^{bb'} \, \delta^{n_{b} n'_{b'}} \, \delta^{d_{b} d'_{b'}} \, .
\end{array}
\right.
\label{norm_wave_Psi_star_nu}
\end{equation}
The conditions on the 5D fields $\Psi_{L/R}$ become conditions on the KK-wave functions $f_{L/R}^{(b, \, n_b, \, d_b)}$ by using Eq.~\eqref{KK_Psi_star_1}. There is a new Kirchhoff junction condition on the $J/V$-brane from Eq.~\eqref{Kir_junc_mix_Psi}:
\begin{equation}
\left\{
\begin{array}{rcl}
\displaystyle{\sum_{i=1}^N f_L^{(b, \, n_b, \, d_b)} (0, i) = a^{(b, \, n_b, \, d_b)} \, M} & \text{for} & \mathcal{K}_N = \mathcal{S}_N \, ,   \\
\displaystyle{\sum_{i=1}^N \left[ f_L^{(b, \, n_b, \, d_b)} (y, i) \right]_{y=0}^{\ell} = a^{(b, \, n_b, \, d_b)} \, M } & \text{for} & \mathcal{K}_N = \mathcal{R}_N \, .
\end{array}
\right.
\label{JC_nu_Psi_wave_1}
\end{equation}
Moreover, Eq.~\eqref{nu_ELE_5D}, with Eqs.~\eqref{Dirac_KK-Psi_star}, \eqref{KK_Psi_star_1} and \eqref{nu_L_sep}, gives:
\begin{equation}
a^{(b, \, n_b, \, d_b)} \, m_\psi^{(b, \, n_b)} = M \, f_R^{(b, \, n_b, \, d_b)}(0, i) \, .
\label{eq_a}
\end{equation}
For $m_\psi^{(b, \, n_b)} \neq 0$, Eqs.~\eqref{JC_nu_Psi_wave_1} and \eqref{eq_a} together lead to:
\begin{equation}
\left\{
\begin{array}{rcl}
\displaystyle{\sum_{i=1}^N f_L^{(b, \, n_b, \, d_b)} (0, i) = \dfrac{M^2}{m_\psi^{(b, \, n_b)}} \, f_R^{(b, \, n_b, \, d_b)}(0, i)} & \text{for} & \mathcal{K}_N = \mathcal{S}_N \, ,   \\
\displaystyle{\sum_{i=1}^N \left[ f_L^{(b, \, n_b, \, d_b)} (y, i) \right]_{y=0}^{\ell} = \dfrac{M^2}{m_\psi^{(b, \, n_b)}} \, f_R^{(b, \, n_b, \, d_b)}(0, i) } & \text{for} & \mathcal{K}_N = \mathcal{R}_N \, .
\end{array}
\right.
\label{JC_nu_Psi_wave_2}
\end{equation}

\subsubsection{Kaluza-Klein Mode Analysis on the Star Graph}
We give here only the KK-mode analysis of the $N$-star $\mathcal{S}_N$, since the $N$-rose $\mathcal{R}_N$ compactification should be incompatible with cosmology without additional assumptions. For completeness, we give the KK-mode analysis on $\mathcal{R}_N$ in Appendix~\ref{neutrino_rose_app}.\\

There are no zero-modes ($b=0$, $n_0=0$, $m_\psi^{(0, \, 0)}=0$) with $M \neq 0$ for $\mathcal{K}_N = \mathcal{S}_N$. Let us look at massive KK-modes ($m_\psi^{(b, \, n_b)} \neq 0$). The coupled first order differential equations \eqref{wave_eq_Psi_star} can be decoupled into second order ones as in Eq.~\eqref{eq_Psi_wave_2nd}. The KK-wave functions $f_{R}^{(b, \, n_b, \, d_b)}$ are still continuous across the junction. We will use the same method as in Subsections~\ref{KK_scalar_20} and \ref{KK_Drac_20}. We give the results in what follows.

\subsubsection*{\boldmath \textcolor{black}{\textit{First case: $f_R^{(b, \, n_b, \, d_b)}(0,i)=0$}}}
The results are identical to the ones in Subsection~\ref{excited_modes_fermion}, Paragraph ``First case: $f_R^{(b, \, n_b, \, d_b)}(0, i)=0$'' of a) p.~\pageref{1st_case_psi_star}. This condition gives the mass spectrum \eqref{mass_spect_Psi_star_2} which defines the KK-tower $b=1$. We have $a^{(1, \, n_1, \, d_1)} = 0$ from Eq.~\eqref{eq_a} so the left-handed modes do not mix with $\nu_L$: they are completely sterile, interact only with gravity, and are thus part of the hidden sector of the model.

\subsubsection*{\boldmath \textcolor{black}{\textit{Second case: $f_R^{(b, \, n_b, \, d_b)}(0,i) \neq 0$}}}
The KK-mass spectrum is given by the transcendental equation:
\begin{equation}
m_\psi^{(2, \, n_2)} \, \tan \left[ m_\psi^{(2, \, n_2)} \, \ell \right] = \dfrac{M^2}{N} \, , \ \ \ n_2 \in \mathbb{N} \, ,
\end{equation}
whose solutions $m_\psi^{(2, \, n_2)}$ define the KK-tower $b=2$ and are not degenerate ($d_2 \in \{1\}$). We have $a^{(2, \, n_2, \, 1)} \neq 0$ from Eq.~\eqref{eq_a} so the left-handed modes mix with $\nu_L$. The lightest mode $(2, 0, 1)$ is identified with the neutrino we observe in Nature\footnote{Of course, we observe three generations of neutrinos in Nature and here we consider a toy model with only one generation.}. In the decoupling limit $\ell \rightarrow 0$, the mass of this mode is given by Eq.~\eqref{m_nu_2} in the zero-mode approximation. Indeed, in this limit the excited KK-modes decouple and their mixing with the lightest massive mode $(2, 0, 1)$ goes to zero. The KK-wave functions are
\begin{equation}
\left\{
\begin{array}{rcl}
f_L^{(2, \, n_2, \, 1)} (y, i) &=& - \left[ \dfrac{N\ell}{2} + \dfrac{M^2(2N-1)}{2 \left(\left[m_\psi^{(2, \, n_2)}\right]^2 + \left[ \dfrac{M^2}{N} \right]^2 \right)} \right]^{-1/2} \, \sin \left[ m_\psi^{(2, \, n_2)} \, (y-\ell) \right] \, , \\ \\
f_R^{(2, \, n_2, \, 1)} (y, i) &=& \left[ \dfrac{N\ell}{2} + \dfrac{M^2(2N-1)}{2 \left(\left[m_\psi^{(2, \, n_2)}\right]^2 + \left[ \dfrac{M^2}{N} \right]^2 \right)} \right]^{-1/2} \, \cos \left[ m_\psi^{(2, \, n_2)} \, (y-\ell) \right] \, ,
\end{array}
\right.
\end{equation}
where the $f_L^{(2, \, n_2, \, 1)}$'s are discontinuous across the $J$-brane (except for $\mathcal{K}_N = \mathcal{S}_1$, the interval, where they are taken continuous) and propagate in the whole star graph. This discontinuity is sourced by the brane-localized interaction.\\

In a nutshell, the massive KK-modes have still a mass gap of order $1/\ell$. Only the KK-modes, whose right-handed Weyl spinors have nonvanishing wave functions at the junction without brane-localized Yukawa couplings (cf. Subsection~\ref{KK_Drac_20}), are affected by the addition of the brane-localized SM-left-handed neutrino $\nu_L$ and Higgs field. The KK-masses are shifted and the wave functions of the left-handed Weyl spinors become discontinuous at the junction\footnote{In the literature, it is known that an interaction localized on a brane away from a boundary or at a fixed point of the orbifolds $\mathcal{R}_1/\mathbb{Z}_2$ and $\mathcal{R}_1/(\mathbb{Z}_2 \times \mathbb{Z}_2')$ implies a discontinuity for a 5D fermion field \cite{Bagger:2001qi, Csaki:2003sh, Casagrande:2008hr, Casagrande:2010si, Carena:2012fk, Malm:2013jia, Barcelo:2014kha}.}. This result is easy to understand when the Yukawa interaction with the VEV of the Higgs field is treated as a perturbation: only nonvanishing wave functions at the junction in Subsection~\ref{KK_Drac_20} will have nonvanishing matrix elements. These modes mix with the SM-left-handed neutrino. The other ones are completely sterile and interact only through gravity.

\section{Conclusion \& Perspectives}
\label{conclusion_star_rose}
In this work, we have studied the possibility of compactifying a LED on a star/rose graph with identical leaves/petals. In Section~\ref{spacetime_geom}, we have defined the geometry of the two 5D spacetimes, and we have adapted Kurasov's distribution theory to a star/rose graph. In this way, we have defined a rigorous framework to build a field theory on these geometries.\\

In Sections~\ref{KG_field} and \ref{Dirac_field}, we have worked out the KK-decomposition of a Klein-Gordon and Dirac field, respectively. Our main contributions, compared to the previous articles \cite{Kim:2005aa, Fujimoto:2019fzb}, are discussions concerning the different possibilities for the continuity of the fields at the junction and the impact on the KK-spectrum. In particular, we have pointed out the case of an airtight brane (when the off-shell fields are allowed to be discontinuous), which is equivalent to $N$ disconnected bonds. Moreover, we have studied for the first time the KK-modes of a massless 5D Dirac fermion propagating into the whole star/rose graph. We have also discussed the chirality of the zero-modes. For both bosonic and fermionic massless fields, the KK-scale is given by the inverse of the leaf/petal length/circumference.\\

One can realize a large compactified volume with a high KK-scale for a large number of small leaves/petals. This possibility has been investigated in Section~\ref{ADD_star_rose} in order to lower the gravity scale to the TeV-scale and solve the gauge hierarchy problem under the hypothesis of an exact global $\Sigma_N$ symmetry. Moreover, we have shown that if the SM-fields are localized on the 3-brane at the junction of the star/rose graph, they couple only to few modes of the entire tower of KK-gravitons, even when a UV-brane thickness is taken into account. The couplings of the SM-fields to this KK-tower are suppressed by the large 4D Planck scale instead of the 5D one at the TeV-scale: the KK-gravitons are thus completely invisible at the LHC ($\sqrt{s}=14$~TeV). This result is in sharp contrast to standard ADD models in the literature with compactification on a torus or its orbifolds, where the SM-fields couple to the whole tower of KK-gravitons, which implies couplings suppressed only by the 5D Planck scale (near a TeV) which one can test at the LHC.\\

Besides KK-gravitons, our proposition can still be probed at hadronic colliders through the search for strongly coupled phenomena induced by gravity like QBH's at the LHC ($\sqrt{s}=14$~TeV) or semi-classical black holes at the Future Circular Collider proton-proton (FCC-pp, $\sqrt{s}=100$~TeV). The absence of a theory of Planckian gravity renders it difficult to make precise predictions concerning the production and decay of QBH's or other exotic states near the Planck scale. It is thus delicate to translate the LHC data ($\sqrt{s}=13$~TeV) into constraints on the 5D Planck scale of our model and to estimate the degree of fine-tuning which remains to accommodate an EW-scale at 100 GeV.\\

It should be interesting to study the effect of gravity on the global $\Sigma_N$ symmetry and the impact on the resolution of the gauge hierarchy problem. Without a complete theory of quantum gravity in the UV, it is difficult to give a quantitative answer, which goes well beyond the scope of this article.\\

Finally, in Section~\ref{Dirac_Neutrinos} we have proposed to realize in our scenario a toy model of small Dirac neutrino masses. For that purpose, we have considered only one generation of neutrinos coupled to one gauge singlet fermion in the bulk. The large compactified volume suppresses the 5D Yukawa coupling of order unity, and we have been able to reproduce the good order of magnitude for the mass of the neutrinos, with a model which accommodates also a 5D Planck scale at the TeV-scale. This kind of model was discussed previously only with a toroidal/orbifold compactification, and the adaptation to a star/rose graph is new. The model is realistic only for the compactification on a star graph, since the rose graph predicts a large number of massless left-handed sterile neutrinos incompatible with cosmology. Moreover, we have found that our models have a hidden sector consisting of secluded KK-gravitons and sterile KK-neutrinos, possibly populated during the early Universe by the decays of mini-black holes in the bulk. The Planckion could also be a candidate to dark matter.\\

We also want to discuss some perspectives for future investigations. As a follow-up of the present work, it would be important to study the unitarity constraints on our models, since a low gravity scale is known to need a UV-completion at a scale lower than the higher-dimensional Planck scale in standard ADD models with a toroidal compactification \cite{Atkins:2010re, Antoniadis:2011bi}. It is also important to see how the mechanism to produce small neutrino masses is influenced by adding bulk and brane-localized Majorana mass terms, in the way of Ref.~\cite{Dienes:1998sb}. Moreover, strongly coupled gravity at the TeV-scale may generate dangerous brane-localized higher-dimensional operators inducing proton decay, large Majorana neutrino masses and flavor changing neutral currents (FCNC's) \cite{Antoniadis:1998ig}. Without knowledge of the UV-completion, we cannot know if these operators are naturally suppressed. If this is not the case, the natural value of their Wilson coefficients are suppressed only by the TeV-scale and one has to add new ingredients to the scenarii to forbid them, like gauging some global symmetries of the SM as the baryon and lepton numbers \cite{Perez:2015rza} and other flavor symmetries \cite{Berezhiani:1998wt, ArkaniHamed:1998sj, ArkaniHamed:1999yy}.\\

Beyond the motivations for the present work, we stress that the 5D background geometries that we studied here can be used in general to generate feebly coupled interactions. Indeed, the couplings of the whole KK-tower of a 5D field, coupled to SM-fields localized at the junction of the star/rose graph, are in general suppressed by the square root of the compactified volume. One can easily imagine how it can be used to build consistent models of axions and dark matter with order one 5D couplings. Moreover, a 5D field has KK-modes whose wave functions vanish at the junction where the SM is localized: they are thus good candidates for a hidden sector. The star graph compactification with a small number of leaves can also be used to build models of 5D SM-fields, as the theory is chiral at the level of zero-modes for 5D fermions. Generating $N$ fermion zero-modes from only one 5D fermion propagating into a star/rose graph with $N$ leaves/petals and an airtight brane  is interesting from the point of view of flavor physics. Moreover, it would be appealing to see if one can implement a 5D supersymmetric field theory or 5D supergravity on the star/rose background. In every scenario, it is important to investigate different possibilities of field theoretical mechanism to stabilize the leaf/petal length scale.

\acknowledgments
I would like to thank Ulrich Ellwanger for encouragement to develop my ideas of model building with an EDS compactified on a
star/rose graph and for reviewing the manuscript. Thanks to Sylvain
Fichet, Ruifeng Leng, Grégory Moreau, Jérémie Quevillon and Robin Zegers for useful discussions. This research was supported by the IDEX Paris-Saclay, the collège doctoral of the Université Paris-Saclay and the Université Paris-Sud.

\appendix

\section{Conventions}
\label{conventions}

The 5D Minkowski metric is
\begin{equation}
\eta_{MN} = \text{diag}(+1, -1, -1, -1, -1).
\label{metric_1}
\end{equation}
The 4D Dirac matrices are taken in the Weyl representation
\begin{equation}
\gamma^\mu =
\begin{pmatrix}
0 & \sigma^\mu \\
\bar{\sigma}^\mu & 0
\end{pmatrix}
\phantom{000} \text{with} \phantom{000}
\left\{
\begin{array}{r c l}
\sigma^\mu &=& \left( I, \sigma^i \right), \\
\bar{\sigma}^\mu &=& \left( I, -\sigma^i \right).
\end{array}
\right.
\label{gamma_1}
\end{equation}
where $\left( \sigma^i \right)_{i \in\llbracket 1, 3 \rrbracket}$ are the 3 Pauli matrices:
\begin{equation}
\sigma^1 =
\begin{pmatrix}
0 & 1 \\
1 & 0
\end{pmatrix},
\phantom{000}
\sigma^2 =
\begin{pmatrix}
0 & -i \\
i & 0
\end{pmatrix},
\phantom{000}
\sigma^3 =
\begin{pmatrix}
1 & 0 \\
0 & -1
\end{pmatrix},
\end{equation}
and thus a 4D Dirac spinor $\Psi$ can be decomposed into its chiral components
\begin{equation}
\Psi = \Psi_L + \Psi_R
\ \ \ \text{with} \ \ \
\Psi_L \equiv
\begin{pmatrix}
\psi_L \\
0
\end{pmatrix}
\ \ \ \text{and} \ \ \
\Psi_R \equiv
\begin{pmatrix}
0 \\
\psi_R
\end{pmatrix}.
\end{equation}
We have also the 4D chirality operator
\begin{equation}
\gamma^5 = i \prod_{\mu = 0}^3 \gamma^\mu =
\begin{pmatrix}
-I & 0 \\
0 & I
\end{pmatrix},
\label{gamma_2}
\end{equation}
which defines the projectors on 4D chirality
\begin{equation}
P_{L, R} = \dfrac{I \mp \gamma^5}{2},
\label{chirality_projectors_1}
\end{equation}
such that for the 4D Dirac spinor $\Psi$
\begin{equation}
\left\{
\begin{array}{r c l}
\Psi_{L, R} &=& \mp \gamma^5 \, \Psi_{L, R}, \\
\bar{\Psi}_{L, R} &=& \pm \bar{\Psi}_{L, R} \, \gamma^5.
\end{array}
\right.
\end{equation}
With our conventions, the 5D Dirac matrices are
\begin{equation}
\Gamma^M = \left( \gamma^\mu, i \gamma^5 \right).
\label{gamma_3}
\end{equation}

\section{Kaluza-Klein Mode Analysis of a Neutrino Model on the Rose Graph}
\label{neutrino_rose_app}

This appendix refers to the model of Section~\ref{Dirac_Neutrinos}. We give the KK-mode analysis on the $N$-rose $\mathcal{R}_N$ of the exact treatment of Subsection~\ref{exact_treatment}.

\subsection{Zero-Modes}
We are looking for zero-modes ($b=0$, $n_0=0$, $m_\psi^{(0, \, 0)}=0$) with $M \neq 0$ for which the first order differential equations \eqref{wave_eq_Psi_star} are decoupled. For $\mathcal{K}_N = \mathcal{R}_N$, there is no right-handed zero-mode. However, we have $N$ degenerate left-handed zero-modes described by the wave functions of Eq.~\eqref{zero_mode_f_L}: the theory is chiral at the level of the zero-modes. Therefore, the brane-localized mass term generates chirality in the rose graph compactification by lifting the degeneracy between the right and left-handed zero-modes which exists in absence of brane-localized Yukawa couplings (cf. Subsection~\ref{KK_Drac_20}). Moreover, we have $a^{(0, \, 0, \, d_0)} = 0$, which means that the left-handed zero-modes do not mix with the left-handed neutrino localized on the $V$-brane: they are sterile neutrinos which do not participate in neutrino oscillations, and they interact only through gravity. As discussed previously, this scenario with large $N$ is likely to be ruled out by cosmological constraints on the number of light fermion species.

\subsection{Massive Modes}

\subsubsection*{\boldmath \textcolor{black}{\textit{First case}: $f_R^{(b, \, n_b, \, d_b)}(0,i) = 0$}}
The results are identical to the ones in Subsection~\ref{excited_modes_fermion}, Paragraph ``First case: $f_R^{(b, \, n_b, \, d_b)}(0, i)=0$'' of b) p.~\pageref{1st_case_psi_rose}. By following this discussion, we get two different mass spectra \eqref{mass_spect_Psi_rose_3} and \eqref{mass_spect_Psi_rose_bis} which define the KK-towers $b=1, 2$ respectively. We have $a^{(b, \, n_b, \, d_b)} = 0$ from Eq.~\eqref{eq_a} with $b = 1, 2$ so the left-handed modes do not mix with $\nu_L$: they are completely sterile (hidden sector).

\subsubsection*{\boldmath \textcolor{black}{\textit{Second case}: $f_R^{(b, \, n_b, \, d_b)}(0,i) \neq 0$}}
The KK-mass spectrum is given by the transcendental equation:
\begin{equation}
m_\psi^{(3, \, n_3)} \, \tan \left[ m_\psi^{(3, \, n_3)} \, \dfrac{\ell}{2} \right] = \dfrac{M^2}{2N} \, , \ \ \ n_3 \in \mathbb{N} \, ,
\end{equation}
whose solutions $m_\psi^{(3, \, n_3)}$ define the KK-tower $b=3$ and are not degenerate ($d_3 \in \{1\}$). We have $a^{(3, \, n_3, \, 1)} \neq 0$ from Eq.~\eqref{eq_a} so the left-handed modes mix with $\nu_L$. The lightest massive mode $(3, 0, 1)$ is identified with the observed neutrino. In the decoupling limit $\ell \rightarrow 0$, we recover that the mass of this mode is given by Eq.~\eqref{m_nu_2} of the zero-mode approximation. The KK-wave functions are
\begin{equation}
\left\{
\begin{array}{rcl}
f_L^{(3, \, n_3, \, 1)} (y, i) &=& - \left[ \dfrac{N\ell}{2} + \dfrac{M^2(2N-1)}{2 \left( \left[ m_\psi^{(3, \, n_3)}\right]^2 + \left[ \dfrac{M^2}{2N} \right]^2 \right)} \right]^{-1/2} \, \sin \left[ m_\psi^{(3, \, n_3)} \left( y - \dfrac{\ell}{2} \right) \right] \, , \\ \\
f_R^{(3, \, n_3, \, 1)} (y, i) &=& \left[ \dfrac{N\ell}{2} + \dfrac{M^2(2N-1)}{2 \left( \left[ m_\psi^{(3, \, n_3)}\right]^2 + \left[ \dfrac{M^2}{2N} \right]^2 \right)} \right]^{-1/2} \, \cos \left[ m_\psi^{(3, \, n_3)} \left( y - \dfrac{\ell}{2} \right) \right] \, ,
\end{array}
\right.
\end{equation}
where the $f_L^{(3, \, n_3, \, 1)}$'s have a discontinuity across the $V$-brane sourced by the brane-localized interaction. Each mode propagates in the whole rose graph.

\bibliographystyle{JHEP}

\providecommand{\href}[2]{#2}\begingroup\raggedright\endgroup

\end{document}